\documentclass[twocolumn,pre,longbibliography,square,numbers,sort&compress]{revtex4-1}
\usepackage{graphicx,color}
\usepackage{amssymb,amsmath}
\usepackage{ifthen}

\newcommand{\<}{\langle}
\renewcommand{\>}{\rangle}

\newcommand{\beq}{\begin{equation}}
\newcommand{\eeq}{\end{equation}}
\newcommand{\bea}{\begin{eqnarray}}
\newcommand{\eea}{\end{eqnarray}}

\newcommand{\bs}{ \mbox{\boldmath$\sigma$}}

\begin{document}

\title{Are biological systems poised at criticality?}

\author{Thierry Mora$^1$}
\altaffiliation{Present address: Laboratoire de Physique Statistique de l'\'Ecole Normale Sup\'erieure, UMR 8550 of CNRS
associated with Universities Paris 6 et Paris 7, 24 rue Lhomond, 75231 Paris Cedex 05, France}
\author{William Bialek$^{1,2}$}
\affiliation{$^1$Joseph Henry Laboratories of Physics, Lewis--Sigler Institute for Integrative Genomics, and $^2$Princeton Center for Theoretical Science,
Princeton University, Princeton, New Jersey 08544 USA}

\date{\today}

\begin{abstract}
Many of life's most fascinating phenomena emerge from interactions among many elements---many amino acids determine the structure of a single protein, many genes determine the fate of a cell, many neurons are involved in shaping our thoughts and memories.  Physicists have long hoped that these collective behaviors could be described using the ideas and methods of statistical mechanics.  In the past few years, new, larger scale experiments have made it possible to construct statistical mechanics models of biological systems directly from real data.  We review the surprising successes of this ``inverse'' approach, using examples form families of proteins, networks of neurons, and flocks of birds.  Remarkably, in all these cases the models that emerge from the data are poised at a very special point in their parameter space---a critical point.   This suggests there may be some deeper theoretical principle behind the behavior of these diverse systems. 
\end{abstract}

\maketitle

\section{Introduction}

One of the great triumphs of twentieth century science was the identification of the molecular building blocks of life.  From the DNA molecules whose sequence and structure control the flow of genetic information, to the ion channels and receptors whose dynamics govern the flow of information in the brain, these building blocks are, to a remarkable extent, universal, shared among all forms of life on earth.  Despite the importance of this reduction to elementary constituents, most of what we recognize as the phenomena of life are not properties of single molecules, but rather emerge from the interactions among many molecules.  Almost by definition, what we find especially interesting about the behavior of multicellular organisms (like us) emerges from interactions among many cells, and the most striking behaviors of animal (and human) populations are similarly collective.

For decades, physicists have hoped that the emergent, collective phenomena of life could be captured using ideas from statistical mechanics.  The stationary states of biological systems have a subtle structure, neither ``frozen'' into a well ordered crystal, nor chaotic and disordered like a gas.  Further, these states are far from equilibrium, maintained by a constant flow of energy and material through the system.  There is something special about the states corresponding to functional, living systems, but at the same time it cannot be that function depends on a fine tuning of parameters.  Of the many ideas rooted in statistical physics that have been suggested to characterize these states, perhaps the most intriguing---and the most speculative---is the idea of self--organized criticality.

The  theory of self--organized criticiality has its origin in models for inanimate matter (sandpiles, earthquakes, etc.)  \cite{Bak:1987p7979}, but the theory was then extended and adapted to encompass biological systems through the analysis of simple toy models \cite{Bak}.    As an example, simple models for the evolution of interacting species can self--organize to a critical state in which periods of quiescence are interrupted by ``avalanches'' of all sizes \cite{Bak:1993p8117}, which reminds us of the idea of punctuated equilibria in evolution  \cite{Gould:1977p9361}.  Similarly, it was suggested that the brain is in a self--organized critical state, at the boundary between being nearly dead and being fully epileptic \cite{Usher:1995p9451}.  It now seems unlikely that some of the initial ideas were correct (e.g., real sand behaves very differently from the models), but possibility that biological system poise themselves at or near a critical point remains tantalizing.

Despite the enthusiasm for using ideas from statistical physics to think about biological systems,  the connections between the models and the experimentally measurable quantities often has been tenuous.  Even in the case of neural networks, where statistical physics approaches are perhaps best developed \cite{Hopfield:1982p7781,Hopfield:1986p10439,Amit,Hertz}, the  relationship between  the  models and the dynamics of real neurons is somewhat loose.  For the ideas of criticality, it might not be too harsh to suggest that much of what has been done is at the level of metaphor, rather than calculations which could be tested against real data.

In the past decade or so, there has been an important development in the experimental investigation of biological networks, and this  suggests a very different route to the use of ideas from statistical physics.  While it has long been conventional to monitor the activity or state of individual elements in a network, it is now possible to monitor many elements in parallel.  The technologies are specific to each class of systems---large arrays of electrodes recording from many neurons in parallel \cite{Segev:2004p1530,Litke:2004p11405}, high throughput sequencing to probe large ensembles of amino acid sequences \cite{Weinstein:2009p1566}, accurate imaging to track individual animals in large groups \cite{Ballerini:2008p7908,Cavagna:2008p7912,Cavagna:2008p7911,Cavagna:2008p7909,Ballerini:2008p679}---and each measurement of course has its own limitations.  Nonetheless, the availability of these new experiments has led several groups to try constructing statistical physics models directly from the data.  A remarkable feature of these analyses, scattered across many levels of organization, is the appearance of signatures of criticality.  Whereas twenty--five years ago we had a grand theory with little connection to data, we now have many isolated discussions of particular experiments hinting at similar conclusions.  Our goal here is to bring these analyses together, perhaps rekindling the hopes for a more general theory.

\section{Zipf's law and criticality}\label{sec:zipf}

In the usual examples of critical phenomena, there are some natural macroscopic variables with a singular dependence on parameters that we can control experimentally.  A familiar example is that we can identify the liquid/gas critical point by measuring the density of the fluid as a function of temperature and pressure.   It is worth noting that, sometimes, doing experiments that couple to the correct macroscopic variables is difficult, as in the Bishop--Reppy experiments on superfluid helium films \cite{Bishop:1978p11563}.  In many cases one can also identify criticality in purely thermodynamic measurements, as a singularity in the heat capacity as a function of temperature, or through the behavior of the correlation function of fluctuations in some local variable, such as the magnetization in a magnet.

The difficulty in biological systems is that they are not really equilibrium statistical mechanics problems, so there is no guarantee that we can find relevant macroscopic variables, and certainly it is not clear how to change the temperature.  Even if an Ising spin glass is the correct description of a neural network, for example \cite{Schneidman:2006p1273,Tkacik:2009p7901,Tang:2008p5830}, it is not clear how to measure the analog of the magnetic susceptibility.  Nonetheless it may be true that the probability of finding the system in a particular state is governed by a probability distribution that is mathematically equivalent to the Boltzmann distribution for a system poised at a critical point.

Let us denote by $\bs$ the state of a system. Typically, $\bs$ is a multi-dimensional variable $\bs=(\sigma_1,\ldots,\sigma_N)$, where $\sigma_i$ can be a spin, a letter in a word, the spiking activity of a neuron, an amino acid in a peptide chain, or the vector velocity of bird in a flock. Let us then denote by $P(\bs)$ the probability of finding the system in the state $\bs$. One can formally write this probability as a Boltzmann distribution:
\beq
P(\bs)=\frac{1}{Z}e^{-E(\bs)/k_BT},
\eeq
where $k_B$ is Boltzmann's constant and $Z$ the partition function. Without loss of generality we can set the temperature to $k_BT=1$ and $Z$ to 1, which leads to the following definition for the energy:
\beq\label{eq:energy}
E(\bs)=-\log P(\bs).
\eeq

With the availability of large datasets of biological systems, it now seems possible to construct $P(\bs)$ directly from the data, and to take the corresponding energy function $E(\bs)$ seriously as a statistical mechanics problem. In this section we explore the consequences of that idea, by showing the equivalence between Zipf's law of language and the critical properties of the associated statistical mechanics model.

In our modern understanding of critical phenomena in equilibrium systems, a central role is played by power law dependencies.  Indeed, the exponents of these power laws---describing the dependence of correlations on distance, or the divergence of thermodynamic quantities as a function of temperature---are universal, and reflect fundamental features of the underlying field theory that describes the long wavelength behavior of the system.  Self--organized critical systems also exhibit power laws, for example in the distribution of sizes of the avalanches that occur as a sandpile relaxes \cite{Bak:1987p7979}.  Power laws have also been observed empirically in a wide variety of non--equilibrium systems \cite{Newman:2005p4243}, although many of these claims do not  survive a rigorous assessment \cite{Clauset:2009p9921}.   It is also fair to note that, in contrast to the case of equilibrium critical phenomena, the observation of power laws in these more exotic cases has not led to anything like a general theory.  

There is a very old observation of a power law in a biological system, and this is Zipf's law in language \cite{Zipf}, first observed by Auerbach in 1913 \cite{Auerbach}.  In contrast to examples such as avalanches, where power laws describe the dynamics of the system, Zipf's law really refers to the distribution over states of the system, in the same way that the Boltzmann distribution describes the distribution over states of an equilibrium system.  Specifically, in written language we can think of the state of the system as being a single word {\bs}, and as texts or conversations proceed they sample many such states.  If one orders (ranks) words {\bs} by their decreasing frequency $P(\bs)$, Zipf's law states that the frequency of words $P(\bs)$   decays as the inverse of their rank $r(\bs)$:
\beq
P({\bs})\propto \frac{1}{r(\bs)}.
\eeq
This distribution cannot be normalized when the number of words is infinite. This can be corrected either by introducing a cutoff corresponding to a finite vocabulary, or  by slightly modifying the law to $P= r^{-\alpha}/\zeta(\alpha)$, with $\alpha>1$ and $\zeta(\alpha)$ is Riemann's zeta function. Since its introduction in the context of language, Zipf's law has been observed in all branches of science, but has also attracted a lot of  criticism, essentially for the same reasons as other power laws, but also because of the controversial claim by Zipf himself that his law was characteristic of human language.  

Despite all our concerns, Zipf's law is, in a certain precise sense, a signature of criticality  \cite{Stephens:2008p2578}.  To see this, consider the density of states, obtained just by counting the number of states in a small window $\delta E$,
The density of states is the number of states within a small energy bracket:
\beq
\rho_{\delta E}(E)=\frac{1}{\delta E}\sum_{\bs} \mathbb{I}[E<E(\bs)<E+\delta E],
\eeq
where $\mathbb{I}[x]$ is the indicator function.  This density of states is the exponential of the entropy, and in the thermodynamic limit the energy and the entropy both should scale with the system's size $N$:
\beq
S(E)\equiv \log \rho_{\delta E}(E)= Ns(\epsilon=E/N) +  s_1 ,
\eeq
where $s_1$ is sub--extensive, that is  $\lim_{N\rightarrow\infty} (s_1/N) = 0$.  The bin size $\delta E$ only affects the sub--extensive corrections as $\delta E\to 0$, and can be ignored for very large systems. But for real data and finite $N$, the choice of the bin size $\delta E$ can be problematic, and it is useful to consider instead the cumulative density of states:
\beq
\mathcal{N}(E)=\sum_{\bs} \mathbb{I}[E(\bs)<E]=\int_{-\infty}^{E}d E' \rho_{\delta E=0}(E').
\eeq
For large systems, this integral is dominated by the maximum of the integrand, and the two definitions for the density of states become equivalent:
\begin{eqnarray}
\mathcal{N}(E)&=&\int_{-\infty}^{E}dE' e^{Ns(E'/N)}\\
&=&N \int_{-\infty}^{E/N}d\epsilon' \exp\left[ N \left( s(\epsilon')+ s_1/N\right)\right]\\
&\sim& e^{Ns(\epsilon)},\\
\Rightarrow \log \mathcal{N}(E)&\sim& Ns(E/N)=S(E).
\end{eqnarray}
But the rank $r(\bs)$ is exactly the cumulative density of states at the energy of $\bs$: $$r(\bs)=\mathcal{N}[E=E(\bs)],$$
that is, the number of states that are more frequent (or of lower energy) than $\bs$, and so in general we expect that, for large systems,
\beq
S[E(\bs)] \approx \log r(\bs)  .
\label{ent-rank}
\eeq

Zipf's law tell us that probabilities are related to ranks,
\begin{eqnarray}
P(\bs ) &=& {1\over{\zeta(\alpha )}} r^{-\alpha}(\bs ) \nonumber\\
\Rightarrow -\log P(\bs ) &=& \alpha \log r(\bs ) + \log \zeta(\alpha ) .
\end{eqnarray}
But now we can connect probabilities to energy,  from Eq (\ref{eq:energy}), and ranks to entropy, from Eq (\ref{ent-rank}), to give
\beq\label{eq:SproptoE}
S(E)=\frac{E}{\alpha}+ \cdots ,
\eeq
where again $\cdots$ is sub--extensive.  In words, Zipf's law for a very large system is equivalent to the statement that the entropy is an exactly linear function of the energy.

A perfectly linear relation between entropy and energy is very unusual.  To see why---and to make the connection to criticality---let's recall the relation of the (canonical) partition function to the energy/entropy relationship.  As usual we have
\beq\label{eq:partfunc}
Z(T)=\sum_{\bs} e^{-E(\bs)/k_BT},
\eeq
where we have reintroduced a fictious temperature $T$. The ``operating temperature,'' {\em i.e.} the temperature of the original distribution, is $k_BT=1$. Then we have
\beq
Z(T) =\int dE \rho(E) e^{-E/k_BT} ,
\eeq
where $\rho(E)$ is the density of states as before.  But in the same large $N$ approximations used above, we can write
\begin{eqnarray}
Z(T)&=&\int dE \rho(E) e^{-E/k_BT} \nonumber\\
&=& \int dE e^{S(E)} e^{-E/k_BT}\\
&\sim& \int d\epsilon \exp\left[ N \left( s(\epsilon) - \epsilon/k_B T\right) \right] .
\end{eqnarray}
For large $N$, this integral is dominated by the largest term of the integrand, which is the point where $ds/d\epsilon = 1/k_B T$; this much is standard, and true for all systems.  But in the special case of Zipf's law, we have $ds/d\epsilon = 1/\alpha$, for all energies.  What this really means is that $k_B T = \alpha$ is a (very!) critical point: 
for any $k_BT<\alpha$, the system freezes into a ground state of zero energy and zero entropy, while for $k_B T > \alpha$ the system explores higher energies with ever higher probabilities, and all thermodynamic quantities diverge if Zipf's law holds exactly.

Clearly, not all critical systems are described by a density of states as restrictive as in Eq \eqref{eq:SproptoE}.
Systems exhibiting a first order transition have at least one energy $E$ for which $S''(E)<0$, and systems with a second order phase transition are characterized by the existence of an energy where $S''(E)=0$. The specific heat, whose divergence serves to detect second order phase transitions, can be related to the second derivative of the micocanonical entropy:
\begin{equation}
C(T) = {N\over {T^2}}\left[ - {{d^2 S(E)}\over{dE^2}}\right]^{-1} .
\end{equation}
What is truly remarkable about Zipf's law, and its correlate Eq \eqref{eq:SproptoE}, is that $S''(E)=0$ {\em at all energies}, making Zipf's law a very strong signature of criticality. A tangible consequence of this peculiar density of states is that the entropy is sub--extensive below the critical point, $S/N\to 0$. For real data, finite size effects will complicate this simple picture, but this argument suggests that critical behaviour can considerably reduce the space of explored states, as measured by the entropy. In later sections, we will see examples of biological data which obey Zipf's law with surprising accuracy, and this observation will turn out to have practical biological consequences.

\section{Maximum entropy models}

Systems with many degrees of freedom have a dauntingly large number of states, which grows exponentially with the system's size, a phonemon sometimes called the `curse of dimensionality'. Because of that, getting a good estimate of $P(\bs)$ from data can be impractical.  The principle of maximum entropy \cite{Jaynes:1957p4009,Jaynes:1957p4011} is a strategy for dealing with this problem by assuming a model that is as random as possible, but that agrees with some average observables of the data.   As we will see, maximum entropy models naturally map onto known statistical physics models, which will ease the study of their critical properties.

In the maximum entropy approach, the real (but unknown) distribution $P_{\rm r}(\bs)$ is approximated by a model distribution $P_{\rm m}(\bs)$ that maximizes the  entropy  \cite{CoverThomas}:
\beq
S[P]=-\sum_{\bs}P(\bs)\log P(\bs),
\eeq
and that satistifies
\beq\label{eq:constraints}
\<\mathcal{O}_a(\bs)\>_{\rm m}=\<\mathcal{O}_a(\bs)\>_{\rm r},
\eeq
where $\mathcal{O}_1,\mathcal{O}_2,\ldots$ are observables of the system, and $\<\cdot\>_{\rm r}$ and $\<\cdot\>_{\rm m}$ are averages taken with $P_{\rm r}$ and $P_{\rm m}$ respectively. The key point is that often average observables $\<\mathcal{O}_a\>_{\rm r}$ can be estimated accurately from the data, even when the whole distribution $P_{\rm r}(\bs)$ cannot. $\mathcal{O}_a$ is typically a moment of one or a few variables, but it can also be a global quantity of the system. Using the technique of Lagrange multipliers, one can write the explicit form of the model distribution:
\beq\label{eq:funcform}
P_{\rm m}(\bs)=\frac{1}{Z}e^{\sum_a\beta_a \mathcal{O}_a(\bs)}.
\eeq
$\beta_1,\beta_2,\ldots$ are the Lagrange multipliers associated to the constraints \eqref{eq:constraints} and constitute the fitting parameters of the model. 
When the maximum entropy model is constrained only by the mean value of the energy, $\mathcal{O}(\bs)=-E(\bs)$, we recover the Boltzmann distribution,
$P_{\rm m}(\bs)=Z^{-1}e^{-\beta E(\bs)}$,
where $\beta=1/k_BT$ is the inverse temperature. More generally, the exponential form of the distribution \eqref{eq:funcform} suggests to define the energy as: $E(\bs)=-\sum_a\beta_a \mathcal{O}_a(\bs)$.

There exists a unique set of Lagrange multipliers that satisfies all the constraints, but finding them is a computationally difficult inverse problem.
Inverse problems in statistical mechanics have a long history, which goes at least as far back as Keller and Zumino, who infered microscopic interaction potentials from thermodynamic quantities \cite{Keller:1959p11443}. The special case of binary variables constrained by pairwise correlations was formulated in 1985 by Ackley, Hinton, and Sejnowski in their discussion of   ``Boltzmann machines'' as models for neural networks \cite{Ackley:1985p10577}. Solving the inverse problem is equivalent to minimizing the Kullback--Leibler divergence between the real and the model distribution \eqref{eq:funcform}, defined as:
\beq
D_{KL}(P_{\rm r}\Vert P_{\rm m})=\sum_{\bs} P_{\rm r}(\bs)\log\frac{P_{\rm r}(\bs)}{P_{\rm m}(\bs)},
\eeq
or equivalently, to maximizing the log-likelihood $\mathcal{L}$ that the experimental data (given by $M$ independent draws $\bs^1,\ldots,\bs^M$) was produced by the model:
\beq
\begin{split}
\mathcal{L}&=\log \prod_{a=1}^M P_{\rm m}(\bs^a)\\
&=M\sum_{\bs} P_{\rm r}(\bs)\log P_{\rm m}(\bs)\\
&=M\left\{S[P_{\rm r}]-D_{KL}(P_{\rm r}\Vert P_{\rm m})\right\}.
\end{split}
\eeq
where, by definition, $P_{\rm r}(\bs)=(1/M)\sum_{a=1}^M \delta_{\bs,\bs^a}$.
In fact, one has:
\beq
\frac{\partial D_{KL}(P_{\rm r}\Vert P_{\rm m})}{\partial \beta_a}=\<\mathcal{O}_a\>_{\rm m}-\<\mathcal{O}_a\>_{\rm r},
\eeq
which ensures that the constraints \eqref{eq:constraints} are satisfied at the minimum.
This explicit expression of the derivatives suggests to use a gradient descent algorithm, with the following update rules for the model parameters:
\beq\label{eq:gradientdescent}
\beta_a\leftarrow \beta_a+\eta(\<\mathcal{O}_a\>_{\rm r}-\<\mathcal{O}_a\>_{\rm m}),
\eeq
where $\eta$ is a small constant, the ``learning rate.''  Note that in this framework, the inverse problem is in fact broken down into two tasks: estimating the mean observables $\<\mathcal{O}_a\>_{\rm m}$ within the model distribution for a given set of parameters $\beta_a$ (direct problem); and implementing an update rule such as \eqref{eq:gradientdescent} that will converge to the right $\beta_a$'s (inverse problem). The direct problem is computationally costly, as it requires to sum over all possible states $\bs$. Approximate methods have been proposed to circumvent this difficulty.
Monte  Carlo algorithms have been commonly used \cite{Tkacik:2009p7901,Shlens:2009p1305,Mora:2010p5398} and have been improved by techniques such as histrogram sampling \cite{Broderick:2007p4970}. Approximate analytic methods, such as high temperature expansions \cite{Sessak:2009p5427,Cocco:2009p2509} or message-passing algorithms \cite{Mezard:2009p2632,Weigt:2009p3341}, were also developed, and shown to be fast and accurate in the perturbative regime of weak correlations.

Note that even when a solution to the inverse problem can be found, one still needs to evaluate whether the maximum entropy distribution correctly describes the data, for example by testing its predictions on local and global observables that were not constrained by the model.   In the following two sections we present examples in which maximum entropy models were successfully fitted to real biological data, and analyzed to reveal their critical properties.  We then turn to other approaches that also point to the criticality of different biological systems.

\section{Networks of neurons}

Throughout the nervous systems of almost all animals, neurons communicate with one another through discrete, stereotyped electrical pulses called action potentials or spikes \cite{Spikes}.   Thus, if we look in a brief window of time $\Delta\tau$,  the activity of a neuron (denoted by $i$) is binary: in this brief window,  a neuron either spikes, in which case we assign it $\sigma_i=1$, or it does not, and then $\sigma_i=-1$.  In this notation the binary string or `spike word' $\bs=(\sigma_1,\ldots,\sigma_N)$ entirely describes the spiking activity of a network of $N$ neurons, and  the probability distribution $P(\bs)$ over all $2^N$ possible spiking states describes the correlation structure of the network, as well as defining the ``vocabulary'' that the network has at its disposal to use in representing sensations, thoughts, memories or actions.

For large networks,  sampling  all $2^N$ words is of course impractical.   For many years, much attention was focused on the behavior of single neurons, and then on pairs.  An important observation is that correlations between any two neurons typically are weak, so that the correlation coefficient between $\sigma_i$ and $\sigma_{j\neq i}$ is on the order of $0.1$ or less.   It is tempting to conclude that, physicists' prejudices notwithstanding, neurons are approximately independent, and there are no interesting collective effects.  As soon as it became possible to record simultaneously from many neurons, however, it became clear that this was wrong, and that, for example, larger groups of neurons spike simultaneously much more frequently than would be expected if spiking were independent in every cell \cite{Meister:1995p10448}.  It is not clear, however, how to interpret such data.  It might be that there are specific sub--circuits in the network that link special groups of many cells, and it is these groups which dominate the patterns of simultaneous spiking.  Alternatively, the network could be statistically homogenous, and simultaneous spiking of many cells could emerge as a collective effect.  An important hint is that while correlations are weak, they are widespread, so that \emph{any} two neurons that plausibly are involved in the same task are equally likely to have a significant correlation.

To make this discussion concrete, it is useful to think about the vertebrate retina.   The retina is an ideal place in which to test ideas about correlated activity, because it is possible to make long and stable recordings of many retinal ganglion cells---the output cells of the retina, whose axons bundle together to form the optic nerve---as they respond to visual stimuli.   In particular, because the retina is approximately flat, one can record from the output layer of cells by placing a piece of the retina on an array of electrodes that have been patterned onto to a glass slide, using conventional methods of microfabrication.    Such experiments routinely allow measurements on $\sim 100$ neurons, in some cases sampling densely from a small region of the retina, so that this represents a significant fraction of all the cells in the area covered by the electrode array \cite{Segev:2004p1530,Litke:2004p11405}.

The average rate at which neuron $i$ generates spikes is given by $\bar r_i = \langle (1+\sigma_i)/2\rangle/\Delta \tau$, so that knowing the average rates is the same as knowing the local magnetizations $\langle\sigma_i\rangle$.  The maximum entropy model consistent with these averages, but with no other constraints,    is a model of independently firing cells, from Eq.~\eqref{eq:funcform}:
\beq
P_1(\bs)=\prod_i p_i(\sigma_i)=Z^{-1}\exp\left[\sum_i h_i \sigma_i\right],
\eeq
where $h_i$ is the Lagrange multiplier associated to the average observable $\<\sigma_i\>$. Although the independent model  may correctly describe the activity of small groups of neurons, it is often inconsistent with some global properties of the network. For example, for the retina stimulated by natural movies \cite{Schneidman:2006p1273}, the distribution of the total number of spikes $K = \sum_{i=1}^N (1+\sigma_i)/2$ is observed to be approximately exponential [$P(K) \approx e^{-K/\bar K}$], while an independent model predicts Gaussian tails.  This suggests that correlations strongly determine the global state of the network.

As the first step beyond an independent model, one can look for the maximum entropy distribution that is consistent not only with $\<\sigma_i\>$, but also with pairwise correlation functions between neurons $\<\sigma_i\sigma_j\>$. The distribution then takes a familiar form:
\beq\label{eq:ising}
 P_2(\bs)=\frac{1}{Z}e^{-E(\bs)},\  E(\bs)=-\sum_{i=1}^N h_i\sigma_i-\sum_{i<j}J_{ij}\sigma_i\sigma_j,
\eeq
where $J_{ij}$ is the Lagrange multiplier associated to $\<\sigma_i\sigma_j\>$. Remarkably, this model is mathematically equivalent to a disordered Ising model, where $h_i$ are external local fields, and $J_{ij}$ exchange couplings. Ising models were first introduced by Hopfield in the context of neural networks to describe associative memory \cite{Hopfield:1982p7781}. The maximum entropy approach allows for a direct connection to experiments, since all the parameters $h_i$ and $J_{ij}$ are determined from data.

\begin{figure}
\begin{center}
\noindent\includegraphics[width=.8\linewidth]{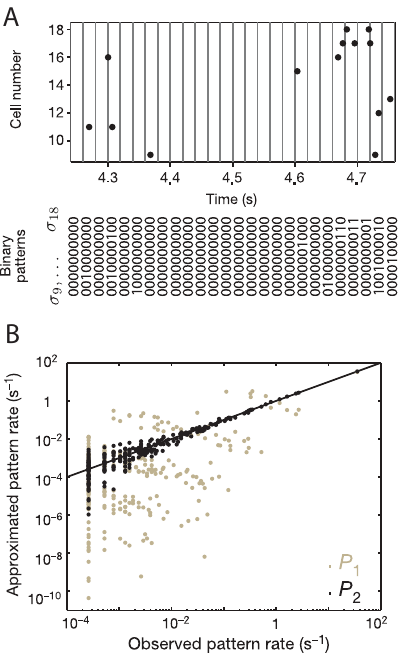}
\end{center}
\caption{The Ising model greatly improves the prediction of retinal activity over the independent model \cite{Schneidman:2006p1273}. {\bf A}. Neuronal activity is summarized by a binary word $\bs=\sigma_1,\ldots,\sigma_N$ obtained by binning spikes into 20 ms windows.
{\bf B}. The frequencies of all spike words $\bs$ of a subnetwork of $N=10$ neurons are compared between the experiment (x axis) and the prediction (y axis) of the independent model (gray dots) and the maximum entropy model with pairwise interactions (black dots). The straight line represents identity.
\label{testmodelschneidman}
}
\end{figure}

Maximum entropy distributions consistent with pairwise correlations, as in Eq \eqref{eq:ising}, were fitted for subnetworks of up to $N=15$ neurons \cite{Schneidman:2006p1273} by direct summation of the partition function coupled with gradient descent [Eq.~\eqref{eq:gradientdescent}].   These models did a surprisingly good job of predicting the  collective firing patterns across the population of all $N$ neurons, as illustrated in Fig.~\ref{testmodelschneidman}.   Importantly, the model of independent neurons makes errors of many orders of magnitude in predicting relative frequencies of the $N-$neuron patterns, despite the fact that pairwise correlations are weak, and these errors are largely corrected by the maximum entropy model.  The accuracy of the model can be further evaluated by asking how much of the correlative structure is captured. The overall strength of correlations in the network is measured by the multi-information \cite{Schneidman:2003p4135}, defined as the total reduction in entropy relative to the independent model, $I=S[P_1]-S[P_{\rm r}]$. The ratio:
\beq
\frac{I_2}{I}=\frac{S[P_1]-S[P_{2}]}{S[P_1]-S[P_{\rm r}]}
\eeq
thus gives the fraction of the correlations captured by the model. When $N$ is small enough ($\leq 10$), $S[P_{\rm r}]$ can be evaluated by directly estimating $P_{\rm r}(\bs)$ from data. In the salamander retina $I_2/I\approx 90\%$, indicating excellent performance of the model.

The generality of the maximum entropy approach suggests that its validity should extend beyond the special case of the salamander retina, and much subsequent work has been devoted to testing it in other contexts.
In a effort parallel to \cite{Schneidman:2006p1273}, the activity of the retina of macaque monkeys \cite{Shlens:2006p1442} was analyzed with maximum entropy methods. The  behaviour of small populations ($N=3$ to $7$) of ON and OFF parasol cells was acurately explained by an Ising model, with 98 to 99\% of the correlations captured. Mammalian retinal ganglion cells can be classified into well-defined types, and cells of a given type tile the visual space like a mosaic \cite{Wassle:1983p10581}; this stands in contrast to the salamander retina, where cells are not well typed and are grouped in large patches responding to the same area of the visual space. It was found that restricting interactions to adjacent pairs in the mosaic did not significantly alter the performance of the model, at least under a limited set of stimulus conditions, a result later confirmed for larger networks   \cite{Shlens:2009p1305}.

The maximum entropy framework was also extended to other (non retinal) areas of the brain. In cultured cortical neurons \cite{Schneidman:2006p1273,Tang:2008p5830} and cortical slices \cite{Tang:2008p5830}, Ising models perfomed as well as in the retina (88 to 95\% of the correlation captured).  Ising models also proved useful for studying neural activity in the visual cortex of cats \cite{Yu:2008p10607} and macaque monkeys \cite{Ohiorhenuan:2010p10598,Ohiorhenuan:2010p10597}.
In monkeys, the Ising model agreed well with data when neurons were far apart from each other ($>600\, \mu$m, tens of micro-columns), but failed at shorter separations ($<300\, \mu$m, a few micro-columns), where higher order correlations prevail \cite{Ohiorhenuan:2010p10597}. This emphasizes the importance of testing the model predictions systematically on local as well as global observables, and if necessary add constraints to the model.

Most of the work reviewed so far was restricted to small population sizes, partly because of the difficulty of recording from many neurons simultaneously, but also because of the computational problems mentioned in the previous section. In the salamander retina \cite{Schneidman:2006p1273}, extrapolations from small networks ($N\leq 15$) have suggested that the constraints imposed by pairwise correlations considerably limit the space of possible patterns (measured by the entropy) as $N$ grows, effectively confining it to a few highly correlated states when $N\approx 200$ --- roughly the size of a patch of retinal ganglion cells with overlapping receptive fields. This led to the proposal that the network might be poised near a critical point.

To test that idea, an Ising model of the whole population of ganglion cells recorded in \cite{Schneidman:2006p1273} ($N=40$) was fitted using Monte Carlo methods and gradient descent \cite{Tkacik:2006p1289,Tkacik:2009p7901}. Although the large size of the population forbids to compute global information theoretic quantities such a $I_2/I$, the validity of the model can still be tested on local observables not fitted by the model. Specifically, the model was found to be a good predictor of the three-point correlation functions $\<\sigma_i\sigma_j\sigma_k\>$ measured in the data, as well as of the distribution of the total number of spikes across the population.

\begin{figure}
\begin{center}
\noindent\includegraphics[width=.8\linewidth]{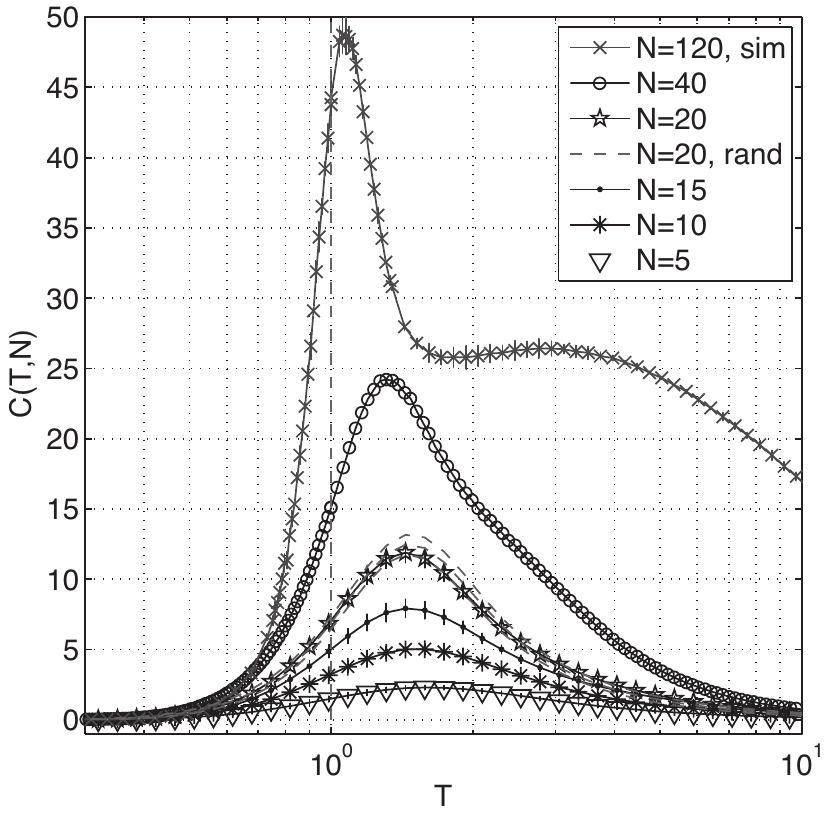}
\end{center}
\caption{Divergence of the heat capacity is a classical signature of criticality. This plot  represents the heat capacity versus temperature for Ising models of retinal activity for increasing population sizes $N$  \cite{Tkacik:2006p1289}. The ``$N=20$, rand,'' and $N=120$ curves were obtained by infering Ising models for fictious networks whose correlations were randomly drawn from real data. Error bars show the standard deviation when choosing different subsets of $N$ neurons among the 40 available.
\label{specheat_retina}
}
\end{figure}

Armed with an explicit model \eqref{eq:ising} for the whole network, one can explore its thermodynamics along the lines sketched in section \ref{sec:zipf}. The introduction of a ficticious temperature $T$ [as in Eq.~\eqref{eq:partfunc}] corresponds to a global rescaling of the fitting parameters, $h_i\to h_i/k_BT$, $J_{ij}\to J_{ij}/T$. As seen in Fig.~\ref{specheat_retina}, the heat capacity versus temperature is found to be more and more sharply peaked around the operating temperature $k_BT=1$ as one increases the network size $N$.   One can also use these ``thermodynamic'' measurements to show that the observed networks of $N\leq 40$ cells are very similar to networks that are generated by mean spike probabilities and correlations at random from the observed distributions of these quantities.    This raises the possibility that criticality could be diagnosed directly from the distribution of pairwise correlations, rather than their precise arrangement across cells.  More concretely, it gives us a path to simulate what we expect to see from larger networks, assuming that the cells that have been recorded from in this experiment are typical of the larger population of cells in the neighborhood.  The result for $N=120$ is an even clearer demonstration that the system is operating near a critical point in its parameter space, as shown by the huge enhancement of the peak in specific heat, shown in the top curve of  Fig.~\ref{specheat_retina}.

\begin{figure}
\begin{center}
\noindent \includegraphics[width=.7\linewidth]{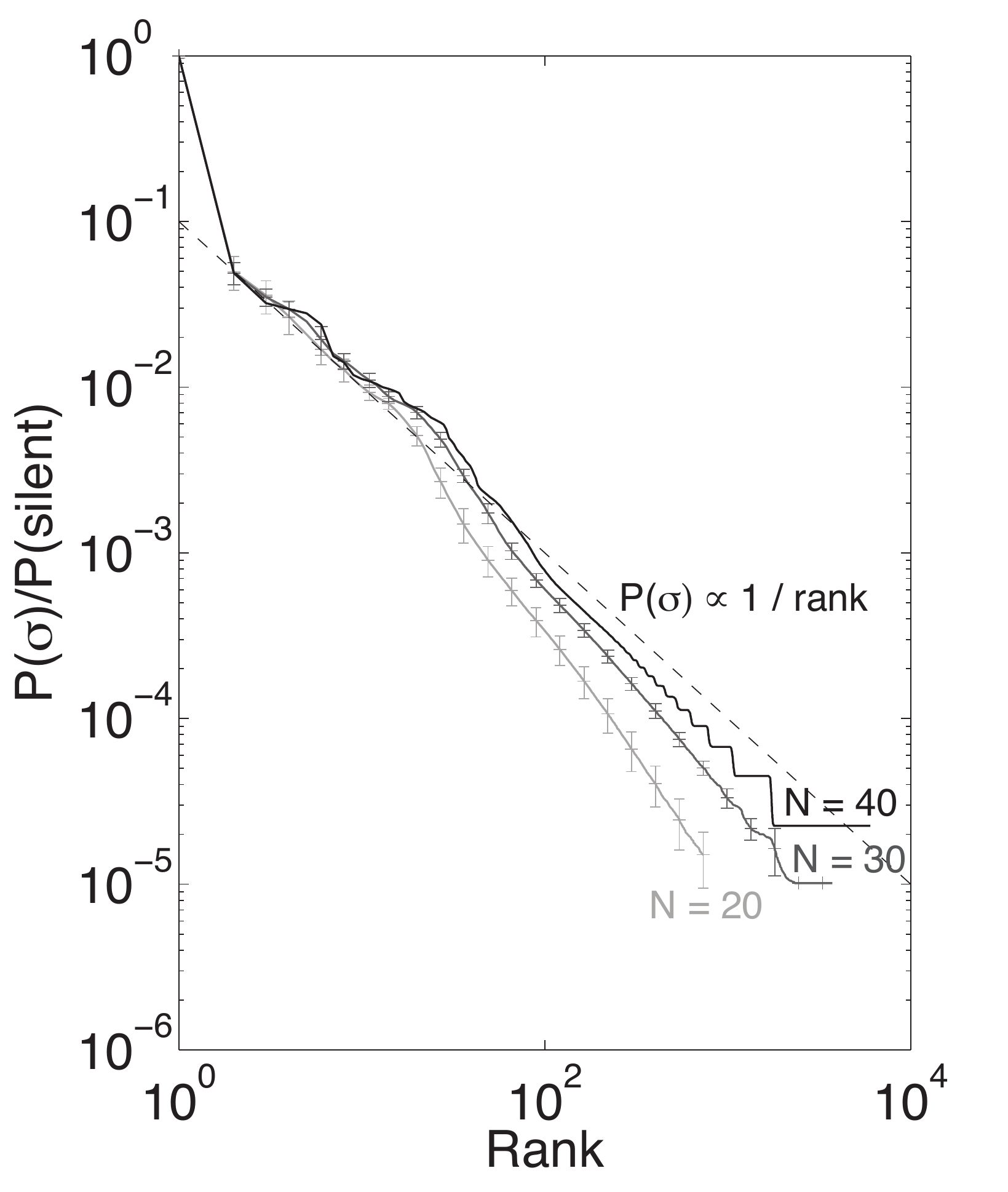}
\end{center}
\caption{The activity of populations of retinal ganglion cells obeys Zipf's law (from the data in Ref \cite{Schneidman:2006p1273}). Shown is the probability of activity patterns (or `words') against their rank for various population sizes. Error bars show the variability across different choices of subpopulations. Note that the agreement with Zipf's law, $P(\bs)\propto 1/$rank, is maximum for larger $N$.
\label{zipf_retina}
}
\end{figure}

This diverging heat capacity is further evidence that the system is near a critical point, but one might be worried that this is an artifact of the model or of the fitting procedure. As we have seen in section \ref{sec:zipf}, the critical properties of the distribution $P(\bs)$ can be also explored directly, without recourse to the maximum entropy approximation, by plotting the probability of firing patterns versus their rank. Figure \ref{zipf_retina}, which shows such plots for increasing network sizes, reveals good agreement with Zipf's law, especially for larger $N$.

Some of the inferred couplings $J_{ij}$ were negative, indicating an effective mutual inhibition between two cells. We know from spin glass theory \cite{Mezard} that negative couplings can lead to frustration and the emergence of many locally stable, or metastable, states. Formally, a metastable state is defined as a state whose energy is lower than any of its adjacent states, where adjacency is defined by single spin flips. Said differently, metastable states are local ``peaks'' in the probability landscape.  In the retina responding to natural movies, up to four metastable states were reported in the population ($N=40$). These states appeared at precise times of the repeated movie \cite{Tkacik:2009p7901}, suggesting that they might code for specific stimulus features. The synthetic network of  $N=120$ cells  displayed a much larger number of metastable states, and the distribution over the basins corresponding to these states also  followed Zipf's law.   At this point however, the exact relation between the proliferation of metastable states and criticality is still not well understood.

In summary, these analyses give strong support to the idea that neural networks might be poised near a critical state.  However, it is still not clear whether the observed signatures of criticality will hold for larger $N$, especially when it is of the order of a correlated patch ($\sim 200$). The next generation of retinal experiments, which will record from $\approx 100-200$ cells simultaneously, should be able to settle that question.

\section{Ensembles of sequences}

The structure and function of proteins is determined by their amino acid sequence, but we have made relatively little progress in understanding the nature of this mapping; indeed, to solve this problem completely would be equivalent to solving the protein folding problem  \cite{Anfinsen:1973p10774,Cordes:1996p5095,Daggett:2003p10775}. An oblique way to tackle that question is to remark that a single function or structure often is realized by many different protein sequences. Can we use the statistics of these related proteins to understand how physical interactions constrain sequences through selection?

To make progress, one first needs to define protein families. Since only a fraction of known proteins have a resolved structure or identified function, defining these families must rely on simplifying assumptions.
The standard method  for constructing a family is to start from a few well identified proteins or protein domains with a common structure or function  \cite{Durbin}. A hidden Markov model is then inferred from that small pool of sequences, and used to scan huge protein databases to search for new members. Clearly, this method only works if the model can set a sharp boundary between members and non--members, and an implicit hypothesis underlying the whole approach is that families are indeed well separated from each other.  

\begin{figure}
\begin{center}
\noindent \includegraphics[width=.8\linewidth]{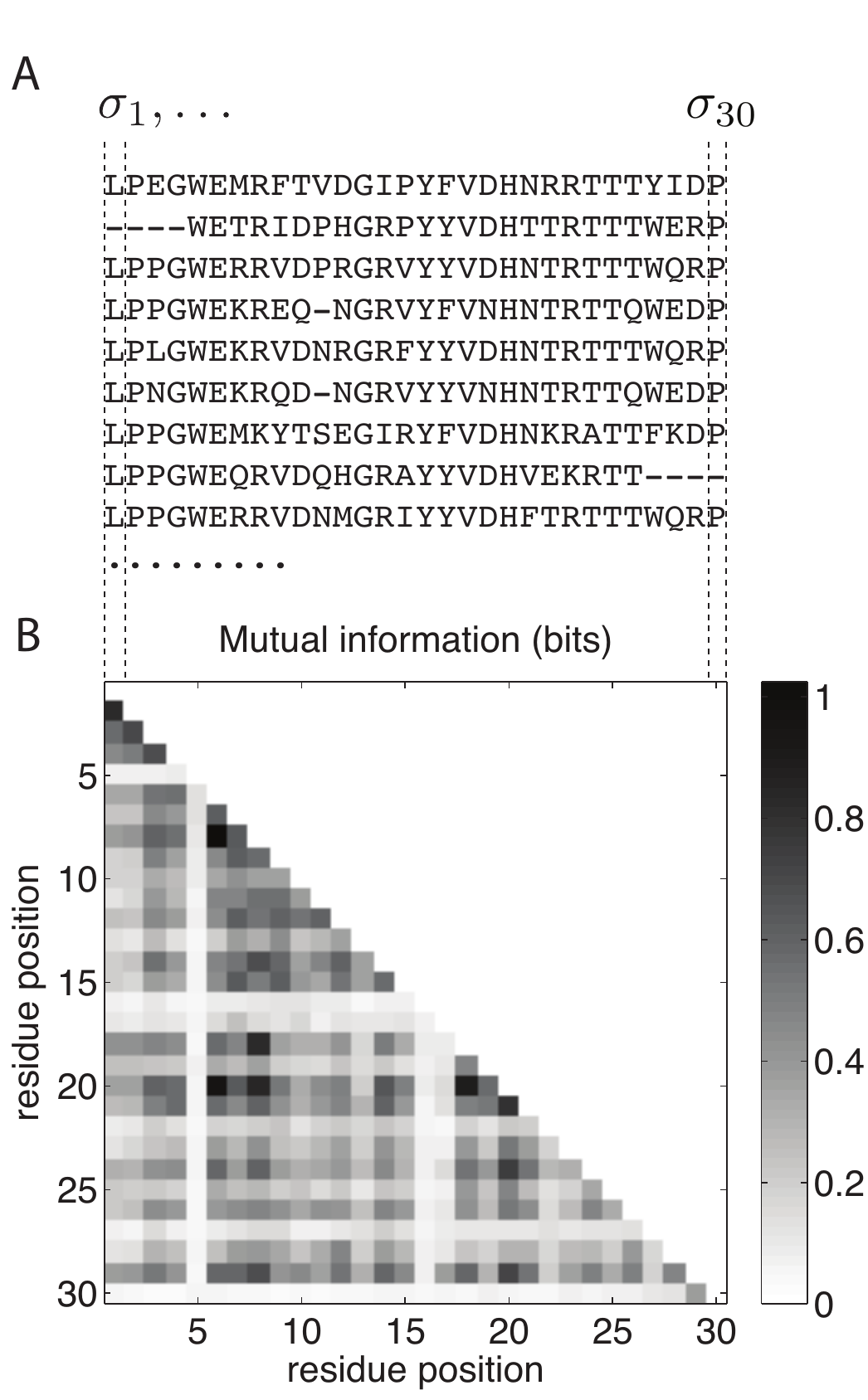}
\end{center}
\caption{Network of correlations between residue positions in the protein family of WW domains.
{\bf A}. A protein sequence is a string $\bs$ of amino acids in the multiple sequence alignment of the family. Here is shown a small sample of co-aligned sequences. {\bf B}. The mutual information between amino acid positions reveals a tighly connected network of correlations between residues all across the sequence.
\label{ww}
}
\end{figure}

Once a protein family has been defined, it is interesting to study its statistical properties.   The data on a particular family consists of a multiple sequence alignment, so that for each member of the family we have a string $\bs=(\sigma_1,\ldots,\sigma_N)$, where $N$ is the number of amino acids in the protein and $\sigma_i$ is one of the 20 possible amino acids at position $i$ in the alignment, or alternatively an alignment gap `--'; cf. Fig.~\ref{ww}A. It is useful to think of the family as a probabilistic object, described by a distribution $P(\bs)$ from which sequences are drawn.   As for networks of neurons, sampling $P(\bs)$ exhaustively is impossible,   so one must have recourse to approximations.

Models of independent residues, $P_1(\bs)=\prod_{i=1}^Np_i(\sigma_i)$, have been widely used in the literature. Physically, however, residues do not simply contribute to the free energy additively \cite{Horovitz:1992p10786}, emphasizing the importance of correlations. Indeed, statistical analyses of protein families reveal strong correlations among the amino acid substitutions at different residue positions in short protein domains \cite{Lockless:1999p10787}. To illustrate this, we represent in Fig.~\ref{ww}B the mutual information between all pairs of positions in a multiple sequence alignment the ``WW domain'' family of proteins.  WW domains are 30 amino acid long protein regions present in many unrelated proteins. They fold as stable, triple stranded beta sheets and bind proline rich peptide motifs. The mutual information gives a measure of correlations between non--numerical variables---here, the residue identity at given positions---defined by
\beq\label{eq:MI}
{\rm MI}[p_{ij}]=\sum_{\sigma_i,\sigma_j}p_{ij}(\sigma_i,\sigma_j)\log_2 \left[\frac{p_{ij}(\sigma_i,\sigma_j)}{p_i(\sigma_i)p_j(\sigma_j)}\right]
\eeq
for a pair of positions $i$ and $j$ in the alignment, where
\bea
p_i(\sigma_i)&=&\sum_{\{\sigma_{k}\}_{k\neq i}}P(\bs),\\
p_{ij}(\sigma_i,\sigma_j)&=&\sum_{\{\sigma_{k}\}_{k\neq i,j}}P(\bs),
\eea
are the one and two point marginals of the distribution, respectively. Some pairs have as much as 1 bit of mutual information among them, which means that one residue can inform a binary decision about the other.

How important are these correlations for specifying the fold and function of proteins? In a groundbreaking pair of papers \cite{Socolich:2005p1730,Russ:2005p1728}, Ranganathan and his collaborators showed that random libraries of sequences consistent with pairwise correlations of WW domains reproduced the functional properties of their native counterpart with high frequency. In contrast, sequences that were drawn from an independent distribution failed to fold. Technically, a random library consistent with pairwise correlations was constructed using a simulated annealing procedure. The algorithm started from the native library and randomly permuted residues within columns of the multiple sequence alignment, thereby leaving the one point functions $p_i(\sigma_i)$ unchanged. The Metropolis rejection rate was designed to constrain the two point functions $p_{ij}(\sigma_i,\sigma_j)$: a cost was defined to measure the total difference between the correlation functions of the native and artificial libraries:
\beq
C=\sum_{i,j,\sigma,\sigma'}\left|\log\frac{p^{\rm native}_{ij}(\sigma,\sigma')}{p^{\rm artificial}_{ij}(\sigma,\sigma')}\right|,
\eeq
and moves were accepted with probability $e^{-\Delta C/T}$, where the algorithm temperature $T$ was exponentially cooled to zero until convergence.

In spirit, this procedure seems similar to the maximum entropy principle: random changes make the library as random as possible, but with the constraint that the one and two point functions match those of the native library. That intuition was  formalized in Ref \cite{Bialek:2007p3854}, where the two approaches were shown to be mathematically equivalent. However, to this day no explicit model for the maximum entropy distribution of the WW domains has been constructed.

The results from \cite{Socolich:2005p1730,Russ:2005p1728} generated a lot of interest, and since then several studies have tried to explore the collective properties of proteins using similar ideas. We now review three of these recent efforts \cite{Weigt:2009p3341,Halabi:2009p3278,Mora:2010p5398}.  All of these examples support the utility of maximum entropy methods in drawing meaningful conclusions about sequence families, while the last focuses our attention back on the question of criticality in these ensembles.

``Two component signaling'' is a  ubiquitous system for the detection and transduction of   environmental cues in bacteria. It consists  of a pair of cognate proteins, a sensor histidine kinase (SK) which detects cellular and environmental signals, and a response regulator (RR) to which signal is communicated by SK via the transfer of a phosphoryl group; the activated RR then triggers other biochemical processes in the cell, including in many cases the expression of other proteins. Many different versions of the two component system are present within and across species, with about 10 per genome on average. A natural question about this system is how the specificity of coupling between particular SK and RR proteins is determined, especially when the different family members have so much in common.  To approach this problem, Weigt {\em et al.} studied a large  collection of cognate SK/RR pairs, and built a maximum entropy model for the (joint) variations in sequence \cite{Weigt:2009p3341,Lunt:2010p10610}.      The maximum entropy distribution consistent with two point correlation functions $p_{ij}(\sigma_i,\sigma_j)$ takes the form of a disordered Potts model:
\beq\label{eq:potts}
P(\bs)=\frac{1}{Z}e^{\sum_i h_i(\sigma_i)+\sum_{ij}J_{ij}(\sigma_i,\sigma_j)},
\eeq
with the gauge constraints $\sum_{\sigma}h_i(\sigma)=0$ and $\sum_{\sigma'} J_{ij}(\sigma,\sigma')=\sum_{\sigma}J_{ij}(\sigma,\sigma')=0$.
The distribution was approximately fitted to the data using mean field techniques \cite{Mezard:2009p2632,Lunt:2010p10610}. 

\begin{figure}
\begin{center}
\noindent \includegraphics[width=.8\linewidth]{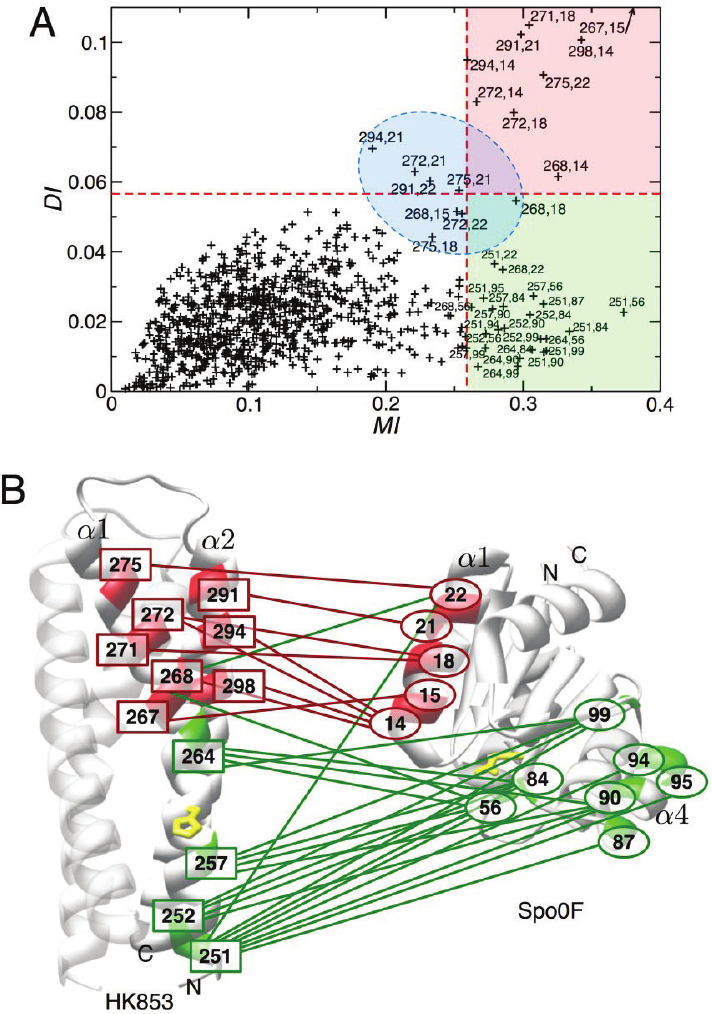}
\end{center}
\caption{The maximum entropy distinguishes between correlations arising from direct pairwise interactions, and correlations arising from collective effects  \cite{Weigt:2009p3341}. {\bf A}. The mutual Information \eqref{eq:MI} between pairs of amino acid position is plotted versus the Direct information \eqref{eq:DI}, which measures the mutual information directly contributed by the pairwise interaction. Among highly correlated pairs, one distinguishes between strongly interacting pairs (red area) and pairs whose correlations result from collective effects (green area) {\bf B}. Direct interactions dominate in the binding domain, while collectively induced correlations are mostly present in the phosphotransfer site.
\label{weigt}
}
\end{figure}

A key point, familiar from statistical mechanics, is that a relatively sparse set of interactions $J_{ij}$ can generate widespread correlations.  It seems plausible that amino acids on the SK and RR proteins which govern the specificity of their contact actually have to {\em interact} in the sense of the Potts model, while other residues may become correlated even if they don't have this essential role in specificity.   The maximum entropy method allows for the distinction of the two cases. A `Direct Information' (DI) was defined as the mutual information between two residues {\em when all other residues are ignored}:
\beq\label{eq:DI}
{\rm DI}_{ij}={\rm MI} [p^{\rm direct}_{ij}]
\eeq
where in
\beq
p^{\rm direct}_{ij}(\sigma,\sigma')=\frac{1}{z_{ij}}e^{J_{ij}(\sigma,\sigma')+h^{(j)}_i(\sigma)+h^{(i)}_j(\sigma')},
\eeq
the `fields' $h^{(j)}_i$ and $h^{(i)}_j$ are chosen such that $\sum_{\sigma}p^{\rm direct}_{ij}(\sigma,\sigma')=p_j(\sigma')$ and $\sum_{\sigma'}p^{\rm direct}_{ij}(\sigma,\sigma')=p_i(\sigma)$. This direct information, which is zero only for $J_{ij}(\cdot,\cdot)=0$, can be viewed as an effective measure of the interaction strength between two residues.
Fig.~\ref{weigt} shows direct information versus mutual information for all pairs of residue positions in the protein complex.
Direct pairwise interactions (large DI, large MI, red) were found to dominate in the binding domain. In contrast, collective effects arising from many weak interactions (low DI, large MI, green) characterized the phosphotransfer domain. Quite naturally, strong interactions (large DI, or equivalently large $J_{ij}$'s) were hypothesized to correspond to direct contact between residues that play a key role in the determination of specificity.

To validate the connection between specificity and the $J_{ij}$, the inferred interacting residue pairs were used to predict the structure of the transient complex formed by the two proteins upon binding. The prediction was shown to agree within crystal resolution accuracy with existing crystallographic data \cite{Schug:2009p4559}.  However efficient, this use of the method only focuses on the strongly interacting pairs involvled in binding, leaving out collective (and possibly critical) behaviors present in the phosphotransfer domain, where strong correlations arise from weak but distributed interactions.  It would be interesting to explore the collective properties of the network as a whole through a more systematic study of the model's thermodynamic properties.

In a parrallel effort, Halabi {\em et al.} \cite{Halabi:2009p3278} showed that variability in protein families could be decomposed into a few collective modes of variations involving non--overlapping groups of residues, called `sectors', which are functionally and historically independent.  To find these sectors, an estimator of the correlation strength was defined as:
\beq
C_{ij}=D_iD_j\left|p_{ij}(\sigma^{\rm cons}_i,\sigma^{\rm cons}_j)-p_{i}(\sigma^{\rm cons}_i)p_j(\sigma^{\rm cons}_j)\right|,
\eeq
where $\bs^{\rm cons}$ is the consensus sequence made of the most common residues at each position. The role of the weights
\beq
D_i=\log\frac{p_i(\sigma^{\rm cons}_i)[1-q(\sigma^{\rm cons}_i)]}
{[1-p_i(\sigma^{\rm cons}_i)]q(\sigma^{\rm cons}_i)},
\eeq
where $q(\sigma)$ is the background probability of residues in all proteins, is to give more importance to highly conserved positions.  The matrix $C_{ij}$ was diagonalized, and the projection of each position $i$ onto the second, third and fourth largest eigenmodes (the first mode being discarded because attributed to historical effets) was represented in a three dimensional space. In that space, which concentrates the main directions of evolutionary variation, residue positions can easily be clustered into a few groups, called sectors.

\begin{figure}
\begin{center}
\noindent \includegraphics[width=.8\linewidth]{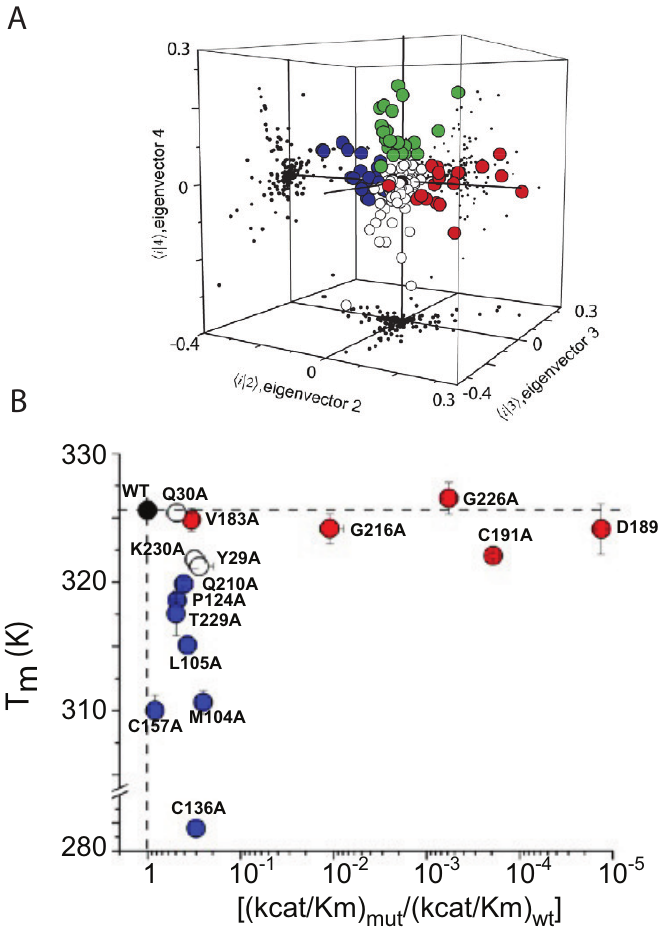}
\end{center}
\caption{Independent sectors in the S1A serine protease family   \cite{Halabi:2009p3278}. {\bf A}. Residue positions $i$ are plotted in the eigenspace of the weighted correlation matrix $C_{ij}$. Three clusters of positions, called sectors, emerge (blue, red and green). {\bf B}. Mutations in different sectors result in independent changes in the biochemical properties of the protein. Mutations in the red sector (red dots) affect the catalytic power ($x$ axis), while mutations in the blue sector (blue dots) change the thermal stability $T_m$ ($y$ axis). 
\label{halabi}
}
\end{figure}

This approach was applied to the S1A serine protease family, for which three sectors were found (Fig.~\ref{halabi}A). Remarkably, two of these sectors are related to two distinct biochemical properties of the protein, namely its thermal stability and catalytic power, and experiments showed that mutations in each sector affected the two properties independently (Fig.~\ref{halabi}B). The mutants used for these experiments were randomly generated by an Ising model identical to \eqref{eq:ising} for each sector. The model was fitted to the data after sequences $\bs$ were simplified to binary strings $\tilde\bs$, with $\tilde\sigma_i=\delta(\sigma_i,\sigma^{\rm cons}_i)$. Although no systematic study of the many body properties of this model was carried out, the non--additive effect of mutations on the protein's properties was demonstrated by experiments on double mutants.

Finally, in recent work, the maximum entropy approach was used to study the diversity of an unambiguously defined family of proteins:  the repertoire of B cell receptors in a single individual \cite{Mora:2010p5398}.   B cells are components of the immune system;  each indivudual has many B cells, each of which expresses its own specific surface receptor (an antibody) whose task is to recognize antigens. Thus, the diversity of the repertoire of B cell receptors carries an important biological function, as it sets the range of pathogens against which the organism can defend itself. The mechanisms by which diversity is generated in the repertoire are complex and not entirely elucidated \cite{Janeway}. Recently, Weinstein {\em et al.} have sequenced almost exhaustively the repertoire of B cell receptors of single zebrafish \cite{Weinstein:2009p1566}, allowing for the first time for a detailed analysis of repertoire diversity.

A main source of the diversity is generated through a process called recombination, which pieces together different segments of the antibody sequence (called V, D and J segments), each of which is encoded in the genome in several versions. Additional diversity is generated at the VD and DJ junctions by random addition and removal of nucleotides during recombination. Finally, antibody sequences undergo random somatic hypermutations, mostly in and around the D segment, throughout the lifetime of the cell. Thus, most of the diversity is concentrated around the D segments, which also constitute one of the three main loops involved in the pathogen recognition process. The D region (defined as the D segment plus its flanking junctions) is therefore a excellent place to study repertoire diversity.

Compared to the previous cases, the definition of the family here is straightforward: all D region sequences of a single individual.  However, and in contrast to other protein families, D sequences cannot be aligned, and their length varies considerably (from 0 to 8 amino acids). To circumvent this problem, a maximum entropy distribution consistent with translation invariant observables was defined. This leads to writing a model similar to Eq \eqref{eq:potts}, but where $h_i=h$ and $J_{ij}=J_{k=(i-j)}$ do not depend on the absolute position of the residues along the sequence. In addition, in order to account for the variable length, the length distribution itself was added to the list of fitted observables, resulting in a chemical potential $\mu[L(\bs)]$ being added to the Potts energy,  where $L(\bs)$ is the sequence length.

\begin{figure}
\begin{center}
\noindent \includegraphics[width=.8\linewidth]{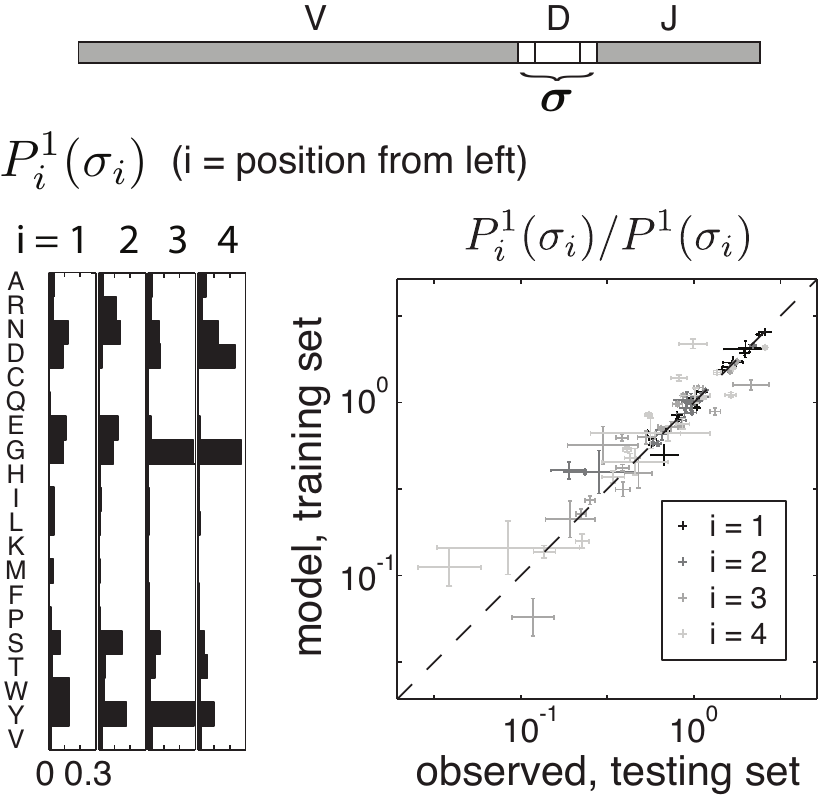}
\end{center}
\caption{A translation invariant maximum entropy model of non-aligned sequences correctly predicts amino acid frequencies at absolute positions   \cite{Mora:2010p5398}. Top: the sequence is made of three segments, V D and J, of which on D and its flanking junctions are fitted by a translation invariant maximum entropy model. Left: for each position $i$ from the left, the frequency table $P^1_i$ for all 20 residues is represented by a histogram. Right: comparison of these frequencies between data and model prediction (after rescaling by the translation-invariant independent model $P^1(\sigma_i)$).
\label{antibody}
}
\end{figure}

The model was fitted by gradient descent combined with Monte Carlo simulations. Pairwise correlation between nearest and second nearest neighbors alone explained 70 to 90\% of correlations, contributing to a large drop in entropy compared to the independent model, from 15 to 9 bits on average. Thus, correlations limited the size of the repertoire by a $\sim 2^6=64$ fold factor. Despite it being translation invariant, the model could also reproduce local observables by simple end effects, such as the $>10\times$ variation in amino acid frequencies at given absolute positions, as shown in Fig.~\ref{antibody}.

\begin{figure}
\begin{center}
\noindent \includegraphics[width=.8\linewidth]{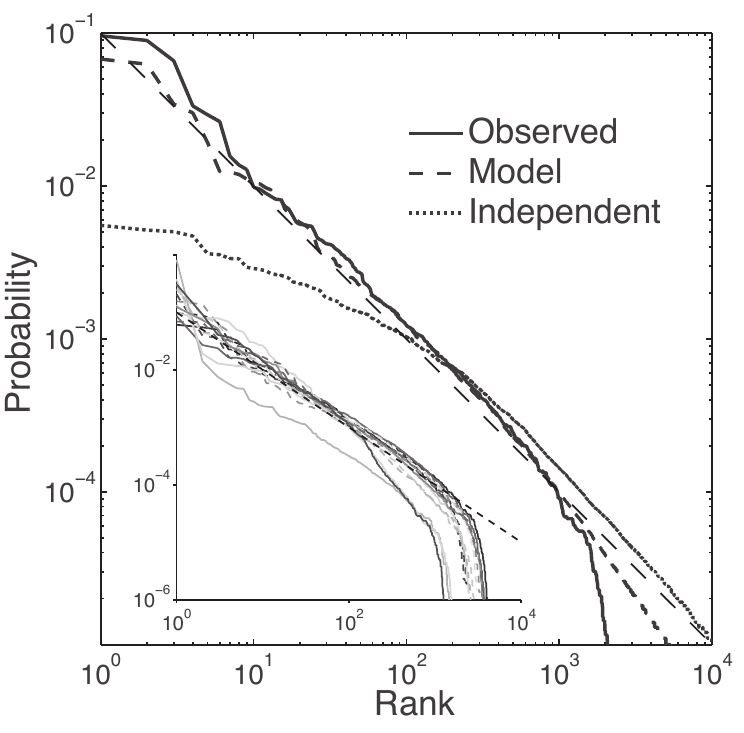}
\end{center}
\caption{The repertoire of antibody D regions of zebrafish follows Zipf's law   \cite{Mora:2010p5398}. For a single fish, the probability of a small antibody segment involved in pathogen recognition is plotted versus its frequency rank, as in Fig.~\ref{zipf_retina}. The data (cyan) is compared with the prediction of a maximum entropy model consistent with nearest and next nearest neighbors correlations (red), and also with a model of independent residues (green). Inset: the same curve plotted for multiple individuals.
\label{zipf_antibodies}
}
\end{figure}

One striking prediction of the model is that the repertoire follows Zipf's law, in close analogy to results obtained for the activity of neural networks. Since the exhaustive sampling of $P(\bs)$ is possible in this case, that prediction can be directly tested against the data, and was found to be in excellent agreement (Fig.~\ref{zipf_antibodies}). Importantly, pairwise correlations between residues are essential for explaining this behavior, as evidenced by the failure of the independent model to reproduce it. The law also seems to be universal, as its varies little from individual to individual, despite substantial differences in the details of their repertoires.

In addition, the model was used to look for metastable states, performing a similar analysis as was done for the retina in the previous section.  About ten relevant metastable states were found for each individual. Not all these states could be mapped onto a genomic template, and it was hypothesized that these non--templated states might reflect the history of antigenic stimulation and thus ``code'' for an efficient defense against future infections. Furthermore, continuous mutation paths existed between almost all metastable states, showing that the repertoire efficiently covers gaps between metastable states, and emphasizing the surprising plasticity of the repertoire.

These results suggest that correlations in protein families build up to create strongly correlated, near--critical states. 
A practical consequence for protein diversity is that collective effects limit the space of functional proteins much more dramatically than previously thought. This should invite us to revisit previously studied families (WW domains,  SK/RR pairs, serine proteases, but also PDZ, SH2, and SH3 domains) to investigate their thermodynamical properties, with the help of maximum entropy models, in search of critical signatures.

\section{Flocks of birds}\label{sec:birds}

Groups of animals such as schooling fish, swarming insects or flocking birds move with fascinating coordination \cite{JensKrause}. Rather than being dictated by a leader or in response to a common stimulus, the collective patterns of flock dynamics tend to be self organized, and arise from local interactions between individuals, which propagate information through the whole group. Flocks, schools and swarms also are highly responsive and cohesive in the face of predatory threat. This balance between order and high susceptibility points to the idea of criticality. Recent field work and theoretical analysis pioneered by the STARFLAG team \cite{Ballerini:2008p7908,Cavagna:2008p7911,Cavagna:2008p7912,Cavagna:2008p7909,Ballerini:2008p679} (see also \cite{Giardina:2008p7914} for a review in relation to previous models), has framed this idea in precise mathematical terms, culminating in the first empirical evidence that flock behaviour may indeed be critical in the sense of statistical physics  \cite{Cavagna:2010p7977}. Before embarking on the description of these results, we first review the technical advances that have made these developments possible.

Three dimensional studies of flocks were pioneered by Cullen {\em et al.} \cite{Cullen:1965p10871}. Until recently, such experiments have focused on small populations of a few tens of individuals, which is insufficient to investigate the large scale properties of flocks. The accurate reconstruction of the three dimensional positions of large flocks is impeded by many technical challenges and has been a major bottleneck. In principle, one can infer the three dimensional coordinates of any object from two photographs taken simultaneously from different viewpoints. But in the presence of a large number of indistinguishable birds, individuals first need to be identified between photographs before that simple geometric argument can be used; this is the so--called matching problem. Use of three cameras can help, but in the presence of noise the matching problem is still highly challenging. In Ref \cite{Cavagna:2008p7911}, new techniques were developed to aid the resolution of the matching problem. The main idea is to compare the patterns formed by the immediate neighborhood of each individual between different photographs. The best match is then chosen as the one maximizing the overlap between these patterns in the different photographs. 

\begin{figure}
\begin{center}
\noindent \includegraphics[width=.95\linewidth]{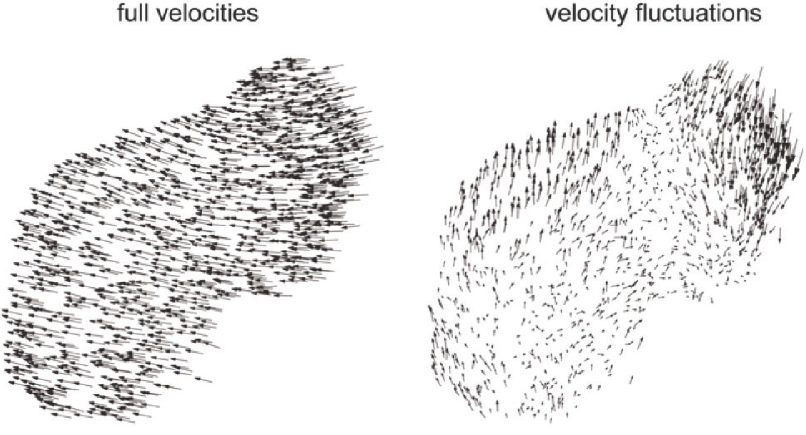}
\end{center}
\caption{Two dimensional projection of a typical 3D reconstruction of the positions and velocities of every bird in a  flock of 1,246 starlings  \cite{Cavagna:2010p7977}. Left: the absolute velocities $\vec v_i$ show a high degree of order in bird orientation. Right: the velocity fluctuations, $\vec u_i=\vec v_i-\frac{1}{N}\sum_{i=1}^N \vec v_i$, are long-ranged, and form only two coherent domains of opposite directions.
\label{flock}
}
\end{figure}

\begin{figure}
\begin{center}
\noindent \includegraphics[width=.95\linewidth]{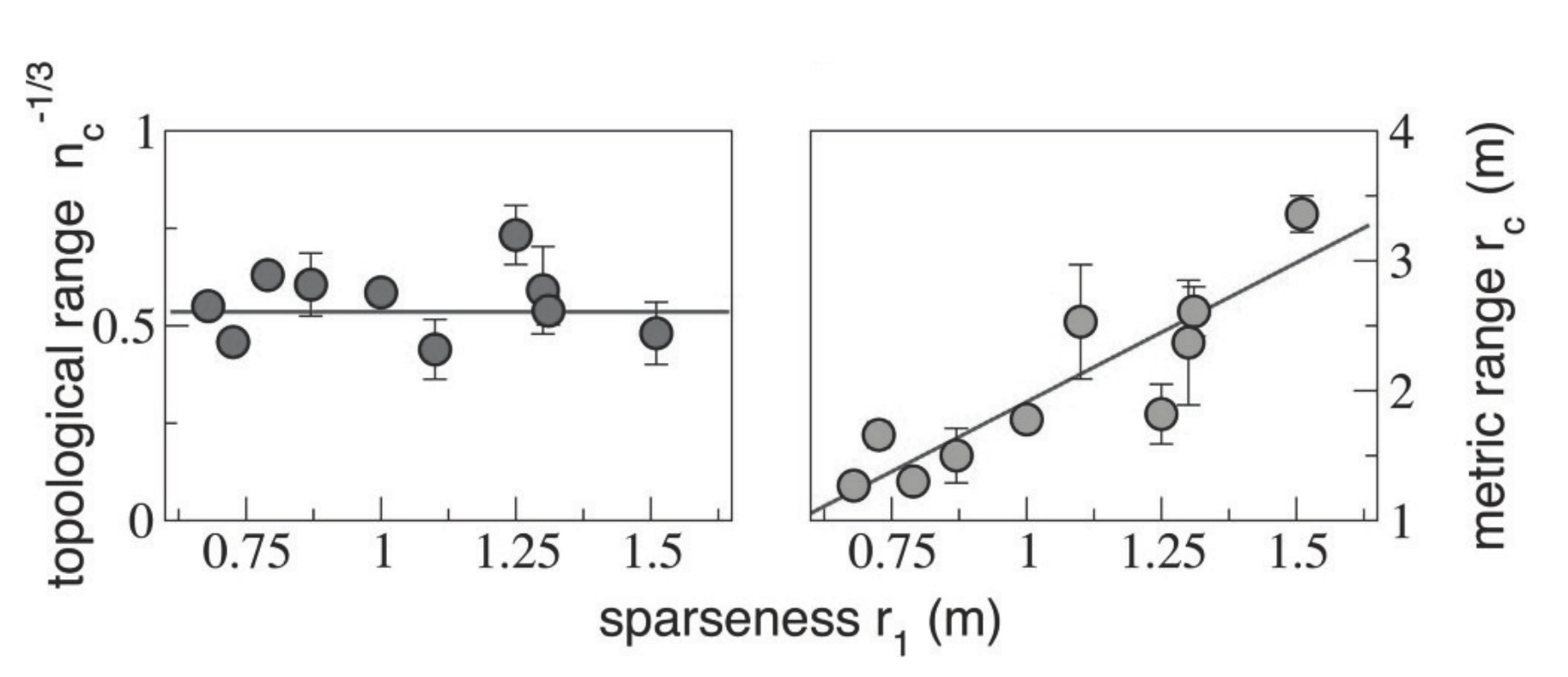}
\end{center}
\caption{Topological versus metric: flocking birds interact with a finite and fixed number of neighbors \cite{Ballerini:2008p679}.  The interaction range is plotted in terms of number of interacting neighbors $n_c$ (left) and in terms of the metric distance $r_c$ (right), as a function of the sparseness $r_1$, defined as the average separation between neighbors. The topological range $n_c \sim 7$ is invariant while the metric range $r_c$ scales with the linear sparseness. 
\label{birdinteraction}
}
\end{figure}

With the help of this technique, triplets of carefully calibrated, high resolution photographs of flocks of starlings taken from three different viewpoints were processed and analysed to yield accurate positions and velocities for all the individuals of flocks comprising up to 2700 birds; see Fig.~\ref{flock} for an example. Preliminary analysis focused on the overall size, shape, density, homogeneity and flying direction of entire flocks \cite{Cavagna:2008p7912,Ballerini:2008p7908}. A subsequent study \cite{Ballerini:2008p679} demonstrated that birds interact with their neighbors according to their topological distance (measured in units of average bird separation), rather than to their metric distance (measured in units of length). The reasoning leading to that conclusion is quite indirect and is worth explaining in some detail. The distribution of neighbors around an average bird is not uniform: birds tend to have closer neighbors on their sides than behind or in front of them. There are biological reasons for this. Birds have lateral vision, and can monitor their lateral neighbors with better accuracy. In addition, keeping a larger distance with frontal neighbors may be a good strategy for avoiding collisions. The main assumption of \cite{Ballerini:2008p679} is that this heterogeneity is a result of interactions between individuals, and can be used to estimate the range of these interactions, defined as the distance at which the neighborhood of an average bird becomes uniform. Plotting this range for various flock densities both in topological and metric units (Fig.~\ref{birdinteraction}) clearly showed that birds interact with a fixed number ($\sim 7$) of neighbors rather than with birds within a fixed radius as was previously thought.

How does global order emerge across the whole flock from local interactions? Clearly, if each bird perfectly mimics its neighbors, then a preferred orientation will propagate without errors through the flock, which will align along that direction. In reality, alignment with neighbors is not perfect, and noise could impede the emergence of global order. This situation is similar to that encountered in physics, where increasing the temperature destroys the ordered state (melting). Consider for example a uniform, fully connected Ising model---the simplest model of ferromagnetism---defined by Eq.~\eqref{eq:ising} with $J_{ij}=J/N$ and $h_{i}=h$. At equilibrium, its mean magnetization $m=\frac{1}{N}\sum_i\<\sigma_i\>=0$ satisfies $m=\tanh(Jm+h)$ \cite{Huang}.  Under a small field $h=0^+$, the system is completely disordered ($m=0$) when the control paramater $J$ (inverse temperature) is smaller than 1, but becomes ordered ($m>0$) for $J>1$.  Interestingly, a similar phase transition occurs in simple models  of flock dynamics \cite{Vicsek:1995p10953}, where the external control parameter can be the noise, the flock density, or the size of the alignment zone. This phase transition, and the concomittent spontaneous symmetry breaking, were analyzed analytically in a continuum dynamical model which exactly reduced to the XY model in the limit of vanishing velocities \cite{Toner:1995p11446,Toner:1998p11445}.

Order is not exclusive to self organized systems, and can instead result from an external forcing (in language appropriate to flocks, by a leader or a shared environmental stimulus). In the Ising model, this corresponds for example to $J=0$ and $h\gg 1$. To better discriminate between 
self-organization and global forcing,
one can examine the response function of the system, or equivalently (by virtue of the fluctuation-dissipation theorem) the correlation functions of small local fluctuations around the ordered state.   In the context of flocks, a large response function means that the flock is not only ordered, but also responds collectively to external perturbations.  It is tempting to suggest that this property is desirable from an evolutionary point of view, as it implies a stronger responsiveness of the group to predatory attacks. We will see that this is indeed how flocks of birds behave. Note that in physical systems, high susceptibility is only achieved near a critical point. In the disordered phase, variables are essentially independent from each other, while in the ordered phase, variables are aligned but their fluctuations become independent as the temperature is lowered.  What is the situation for bird flocks?

\begin{figure}
\begin{center}
\noindent \includegraphics[width=.8\linewidth]{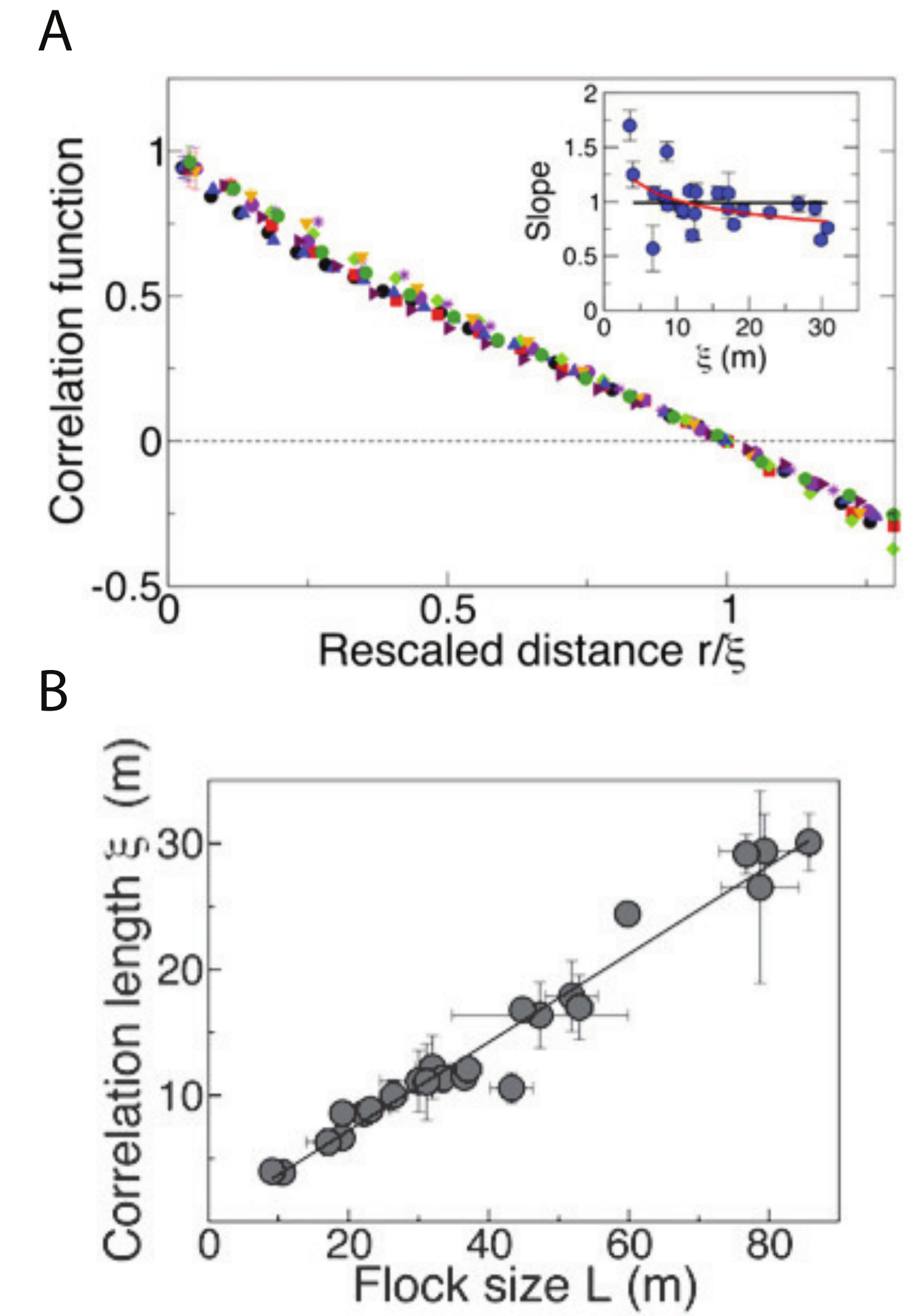}
\end{center}
\caption{Velocity fluctuations are scale free \cite{Cavagna:2010p7977}. {\bf A}.. The correlation length $\xi$ scales linearly with the system's size $L$, indicating that no other scale than $L$ is present in the system.
{\bf B}. Correlation function $C$ versus rescaled distance $r/\xi$. $\xi$ is defined as the radius for which $C=0$. The slope at $r=\xi$ (Inset) seems to depend only weakly upon $\xi$.
This suggests that coherence can in principle be preserved over extremely long ranges.
\label{scalefree}
}
\end{figure}

To explore these ideas empirically, Cavagna {\em et al.} \cite{Cavagna:2010p7977} analyzed the velocity correlations of large flocks, using the same dataset as in previous studies. At this point it should be stressed that here, at variance with the previous cases of neurons and proteins, learning the probability distribution of the system's state is impractical because only one example of the flock's state is available to us. On the other hand, translation invariance (if one excludes the edges of the flock) and homogenpeity in the birds' behavior can be invoked to make statistical statements across the population. Let us call $\vec v_i$ the 3D velocity vector of a bird $i=1,\ldots, N$. The amount of order in the flock is typically measured by the polarization $\Vert\frac{1}{N}\sum_i \frac{v_i}{\Vert\vec v_i\Vert}\Vert$, whose value is here very close to 1 ($0.96\pm0.03$) in agreement with previous studies. But as discussed earlier, more interesting are the fluctuations around the global orientation, defined by the velocities in the reference frame of the center of mass: $\vec u_i=\vec v_i-(1/N)\sum_{i=1}^N\vec v_i$. Correlations in these fluctuations are captured by the distance dependent correlation function:
\beq
C(r)=\frac{1}{c_0}\frac{\sum_{i,j}\vec u_i\cdot\vec u_j \delta(r-r_{ij})}{\sum_{i,j} \delta(r-r_{ij})},
\eeq
where $r_{ij}$ is the distance between birds $i$ and $j$, $\delta(\cdot)$ is a (smoothed) Dirac delta function, and $c_0$ is chosen such that $C(r=0)=1$. The correlation function $C(r)$ is plotted in Fig.~\ref{scalefree}A for different flock sizes as a function of the rescaled distance $r/\xi$, where $\xi$ is a characteristic length defined by $C(\xi)=0$. All points seem to fall onto a single curve.

The results of Fig.~\ref{scalefree}A are consistent with what we know from scaling theory in physics \cite{Huang}. Near a critical point,   correlation functions  are given by a universal function,
\beq\label{eq:scaling}
C(r)=\frac{1}{r^\gamma}f(r/\xi),
\eeq
where $\xi$ is the correlation length which diverges as the critical point is approached.  Strikingly, in bird flocks, the correlation length $\xi$ is found to scale with the linear size of the flock $L$ (Fig.~\ref{scalefree}B). This indicates that the correlation function is in fact scale free, in the sense that no scale is present   except for the system size. Replacing $\xi=\alpha L$ into \eqref{eq:scaling} and taking $L\to\infty$ yields a power law decay for the correlation function, $C(r)={r^{-\gamma}}$, characteristic of a critical point.   The exponent $\gamma$ can in principle be evaluated from data through the derivative of $C$ at $r=\xi$: $\xi{\partial C}/{\partial r}\propto -{\xi^{-\gamma}}$. However, as evident from the inset of Fig.~\ref{scalefree}A, $\gamma$ is almost indistinguishable from zero. This implies that the correlation function is not only scale free, but also decays very slowly, implying extremely strong and long ranged coherence across the flock.

The same analysis was carried out on the correlations of the {\em modulus} of the velocity, rather than its orientation, yielding essentially the same restults. A physical system with a spontaneously broken symmetry, such as its overall orientation, can display scale free (``massless'') behavior of the quantity associated to that symmetry, even when no critical point is present (Goldstone modes). However, the modulus of velocity is a much stiffer mode than velocity orientation, and corresponds to no obvious symmetry. The fact that it also exhibits scale free behavior thus is stronger evidence that the system  indeed is close to a critical point.

One must be cautious when extrapolating from finite system sizes, and conclusions drawn from these extrapolations must be examined with increased scrutiny. Nonetheless, the wealth of evidence in favor of criticality makes it a very useful and pertinent concept for understanding complex flock dynamics.  We expect that continued improvements in experimental technique and data analysis methods will test the hypothesis of criticality much more sharply.

\section{Dynamical vs. statistical criticality}

So far, we have assumed that states of a biological system were drawn from a stationary probability distribution $P(\bs)$, and we have explored questions of criticality in the associated statistical mechanics model. Criticality, however, can also be meant as a dynamical concept. For example, in models of self-organized criticality mentioned in the introduction, avalanches are by nature a dynamical phenomenon \cite{Bak:1987p7979}.     We now discuss two lines of work in this direction: the observation of critical avalanches of activity in networks of cultured neurons,  and dynamical criticality close to a Hopf bifurcation in the auditory system.

We start with avalanches in neural networks \cite{Corral:1995p11377,Herz:1995p8090,Chen:1995p11378}.   Consider a control parameter for neuronal excitability, which sets how much a spike in one neuron excites its neighbors.   If this parameter is too low, a spike in one neuron may propagate to its direct neighbors, but the associated wave of activity will quickly go extinct. Conversely, if the excitability parameter is too high, the wave will explode through the whole population and cause something reminiscent of an epileptic seizure.  To function efficiently, a neural population must therefore poise itself near the critical point between these two regimes. The analogy with sandpiles and earthquakes is straightforward: when a grain falls, it dissipates some its mechanical energy to its neighbors, which may fall in response, provoking an avalanche of events \cite{Bak:1987p7979}. A similar argument applies to earthquakes and the propagation of slips \cite{Bak:1989p11402}.

\begin{figure}
\begin{center}
\noindent \includegraphics[width=.8\linewidth]{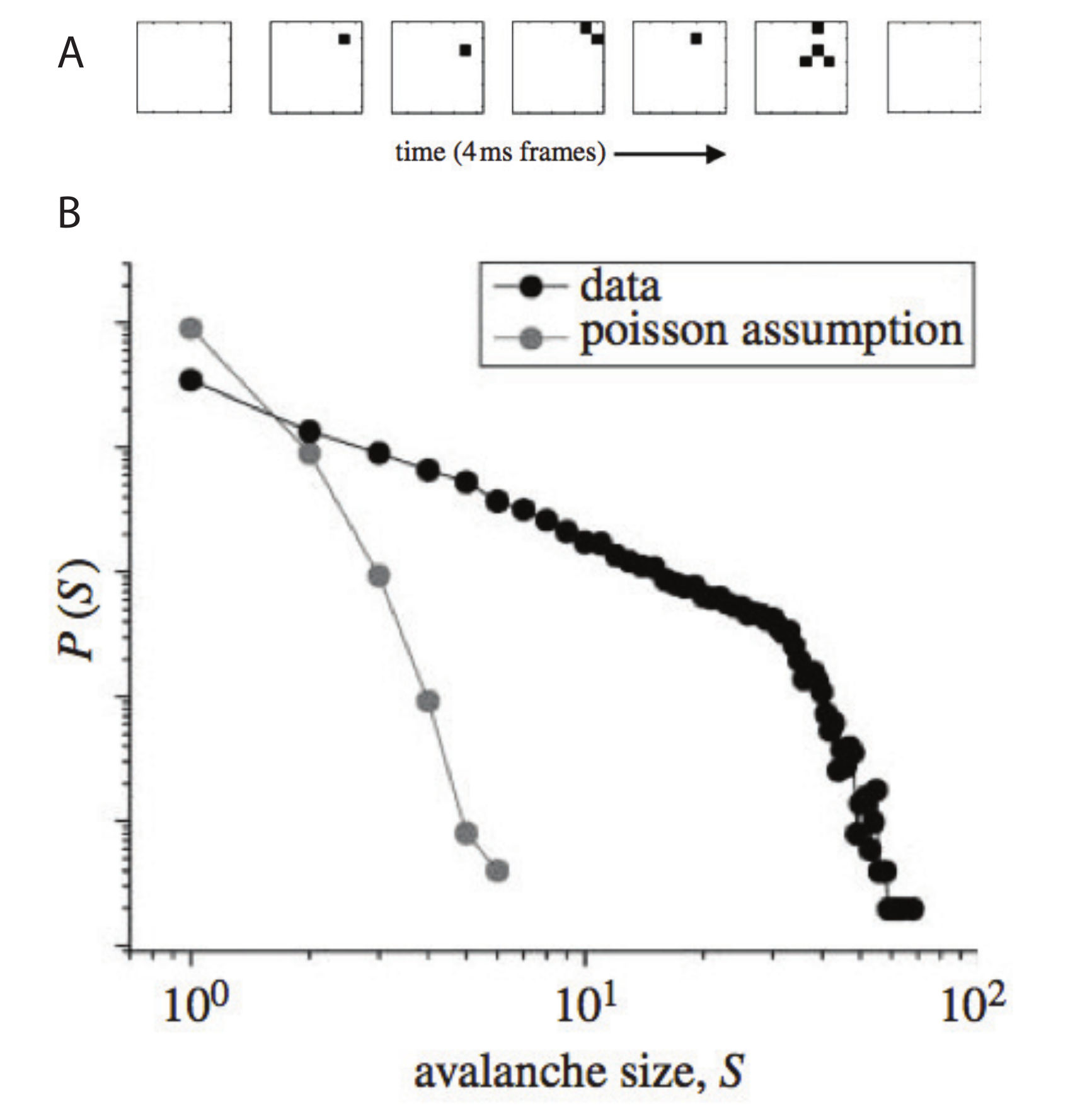}
\end{center}
\caption{The distribution of avalanche sizes follows a power law. {\bf A}. Sample avalanche propagating on the $8\times 8$ multielectrode array. {\bf B}. Probability distribution of avalanche sizes (measured in number of electrode) in log-log space. The distribution follows a  power-law with a cutoff set by the size of the array.
\label{avalanche}
}
\end{figure}

The most striking feature of self--organized criticality is the distribution of the avalanche sizes, which typically follows a power law.  Beggs and Plenz \cite{Beggs:2003p7980} were the first to report such power laws in the context of neural networks.
In their experiment, a 60--channel multielectrode array was used to measure local field potentials (a coarse grained measure of neural activity) in cortical cultures and acute slices. Activity occured in avalanches---bursts of activity lasted for tens of milliseconds and were separated by seconds long silent episodes---that propagated across the array (Fig.~\ref{avalanche}A). For each event, the total number of electrodes involved was counted as a measure of avalanche size. The distribution of this size $s$ followed a power-law with an exponent close to $-3/2$ (Fig~\ref{avalanche}B). Although that exponent was first speculated to be universal, it was later shown that it depends on the details of the measurement method \cite{Beggs:2008p11283}.

The critical properties of neural avalanches can be explained by a simple branching process \cite{Harris}. Assume that when a neuron fires at time $t$, each of its neighbors has a certain probability of firing at time $t+1$, such that the average number of neighbors firing at $t+1$ is given by the branching parameter $\beta$. That parameter is exactly what we called ``excitability'' earlier; $\beta<1$ leads to to an exponential decay of the avalanche, $\beta>1$ to its exponential and unlimited growth, and $\beta=1$ defines the critical point. To support that simple theory, the parameter $\beta$ was estimated directly from the data, and was found to be $1$ within error bars.

\begin{figure}
\begin{center}
\noindent \includegraphics[width=.3\linewidth]{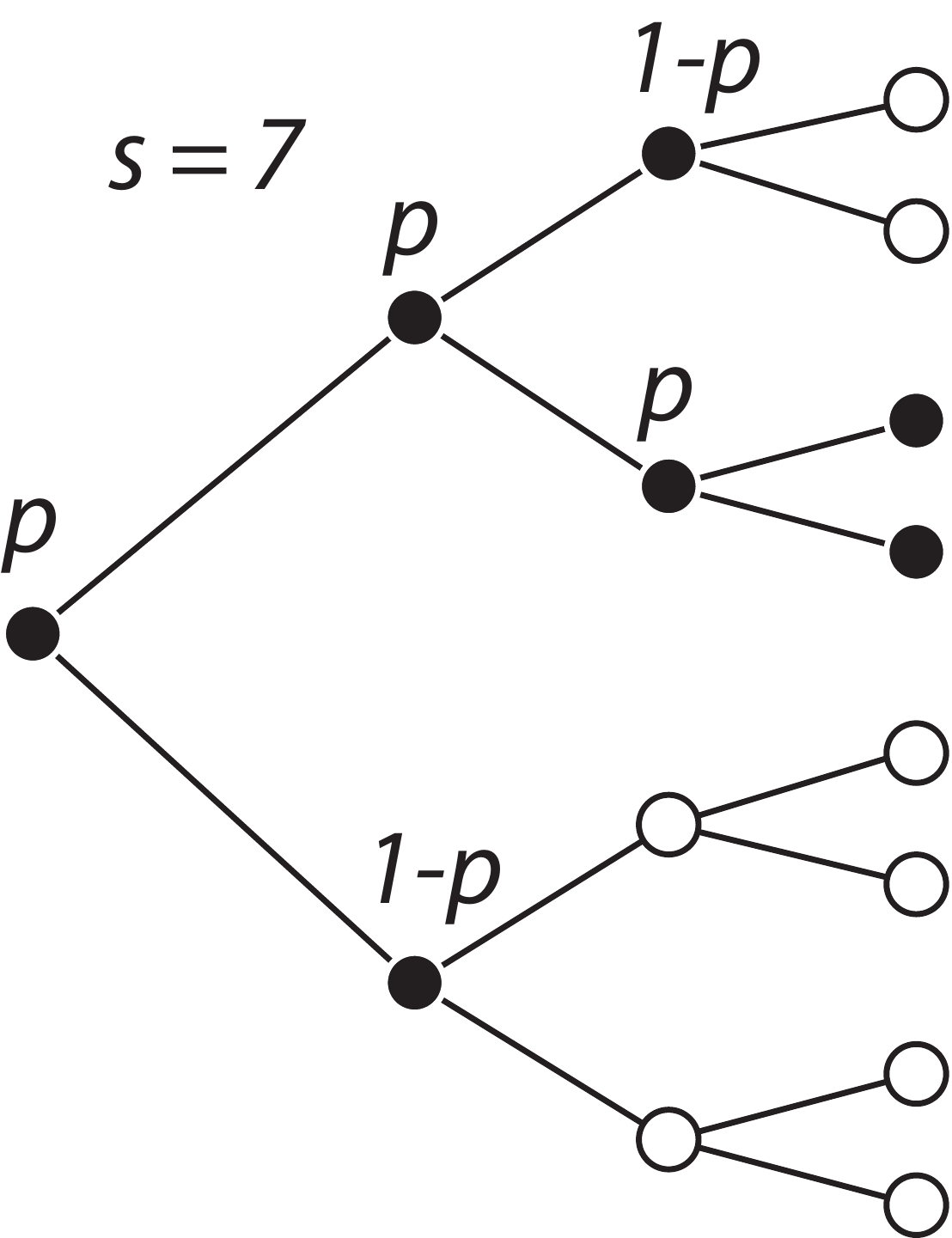}
\end{center}
\caption{A simple branching process on a tree \cite{Zapperi:1995p11270}. Starting from the root, activity propagates to its two descendants with probability $p=1/2$, or to none with probability $1-p$. The process repeats itself for all active descendants. In this example black node are active, while white node are inactive. The size of the avalanche is $s=7$.
\label{branching}
}
\end{figure}

Can we connect this notion of criticality in a neural network to the ideas discussed in Section \ref{sec:zipf}?   Consider, for example, the simple mean field branching process on an infinite tree analyzed in \cite{Zapperi:1995p11270} and summarized by Fig.~\ref{branching}. When $p=1/2$ ($\beta=1$), one can show, by recursively calculating the generating function of the avalanche size $s$, that the distribution of avalanche sizes becomes
\beq\label{eq:ps}
P(s\gg1 )=\sqrt{2/\pi}s^{-3/2}.
\eeq
Although the resemblance of the exponent $3/2$ to that found in \cite{Beggs:2003p7980} is coincidental, this simple process nonetheless predicts a power law in the avalanche size. Similar models defined on lattices or on completely connected graphs were proposed to explore the functional properties of neural avalanches \cite{Chen:2010p8038,Haldeman:2005p8046,Beggs:2003p7980}. When $p=1/2$, the probability of any particular avalanche event $\bs$  is easy to estimate, and is $2^{-s}$, where $s$ is the size of the avalanche; note that there are many states $\bs$ that correspond to the same size $s$.  Using our definition of the ``energy'' from Eq.~\eqref{eq:energy}, we have $E(\bs)=s\log(2)$. By virtue of Eq.~\eqref{eq:ps}, however, in this dynamically critical state the probability that a random configuration has energy $E$ decays less rapidly than an exponential, and this must result from a near perfect balance between energy and entropy:
\beq\label{eq:pe}
P(E)=\frac{1}{Z}e^{S(E)-E}=\frac{\sqrt{2/\pi}}{(\log 2)^{3/2}}E^{-3/2},
\eeq
which implies:
\beq
S(E)=E-\frac{3}{2}\log(E)+\ldots ,
\eeq
and this is (for large $E$) Zipf's law once again.   Note that this result is driven solely by the fact that the distribution of avalanche sizes has a long tail, and not by any specific power  law behaviour.  To summarize, in the space of avalanche configurations we have the same signature of criticality that we have seen in  the retina (Figs.~\ref{specheat_retina} and \ref{zipf_retina}), although in different tissues, with different measurement methods, and assuming different models of activity. This emphasizes the potential generality of Zipf's law and criticality for brain function.

The space of possible avalanches is huge, and one might wonder whether avalanches can serve as a basis for a neural code. In a simple branching process, each avalanche of a given length occurs completly at random and is as probable as any other. But in a real network, with disordered excitabilities and branching parameters, some types of avalanches may be more likely than others, forming attractors  in avalanche space.  Such attractors were detected in the experimental data by clustering all observed avalanche patterns \cite{Beggs:2004p8037}. Remarkably, simulations of disordered branching processes show that a large number of attractors is only possible when the system is close to the critical point (average branching parameter 1) \cite{Haldeman:2005p8046}. These results are reminiscent of those found in the retina, with the difference that attractors are now defined in a dynamical space rather than as metastable states in the space of configurations. As in the retina, the exact function of these attractors for coding is still elusive.

\bigskip

We now turn to another example where dynamical criticality plays an important role, although in a different way, in the context of the auditory system. Our ear is remarkably sensitivive to weak sounds, responding to motions of the same magnitude as thermal noise. As early as 1948, Gold \cite{Gold} proposed that this sensitivity is achieved by compensating damping through an active process. Several observations support this hypothesis.  Hair cells, which convert mechanical movement into electrical current, respond to sounds in a highly non-linear manner, by strongly amplifying low amplitude stimuli at some frequency that is characteristic of each cell. In addition, hair cells display small spontaneous oscillations even in the absence of stimulus. Most dramatically, the ear can actually emit sounds, spontaneously, presumably as the result of damping being (pathologically) over--compensated at some points in the inner ear \cite{Kemp:1978p11571,Zurek:1981}.  

A series of recent works, both theoretical and experimental, have shown that the mechanosensing system of hair cells is tuned close to a Hopf bifurcation, where the system is highly sensitive to stimulation (see \cite{Hudspeth:2010p11135} for a recent review). Before going into the specifics of hair cell biophysics, let us first explain the basic idea.

\begin{figure}
\begin{center}
\noindent \includegraphics[width=.8\linewidth]{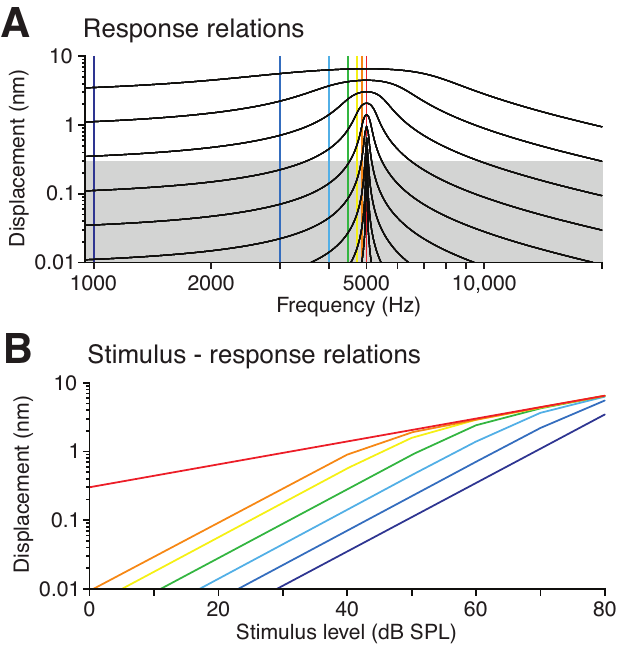}
\end{center}
\caption{Response to an oscillatory input of a Hopf oscillator near its critical point   \cite{Hudspeth:2010p11135}. {\bf A}. Response (displacement) as a function of input frequency, for increasing input amplitudes from 0 dB (lower curve) to 80 dB (top curve). This plot emphasizes the amplification of small inputs, as well as the shrinking width of the  frequency range where amplification is present. {\bf B}. Displacement as a function of simulus amplitude, plotted in log space. The red curve, of slope $1/3$, shows the enhanced response at the critical (resonant) frequency. For other frequencies (whose color correspond to the frequencies marked by lines in A), the response is linear.
\label{hopf}
}
\end{figure}

A Hopf oscillator is described by two essential dynamical variables, often collected into a single complex number $z$. Hopf's oscillators form a universality class of dynamical systems, and in response to small forces near the bifurcation point, the dynamical equations can always be written as
\beq
\frac{dz}{dt}=(\mu+i\omega_0)z-\vert z\vert^2 z+Fe^{i\omega t}.
\eeq
In the absence of forcing, $F=0$, self--sustained oscillations appear for $\mu>0$: $z=re^{i\omega_0 t}$, with $r=\sqrt{\mu}$. When a stimulus is applied at the resonant frequency $\omega_0$, the system simplifies to
\beq
z=re^{i\omega_0 t},\qquad
\label{eq:hopf}
\frac{dr}{dt}=r(\mu-r^2)+F.
\eeq
Precisely at the bifurcation point $\mu=0$, there is no regime of linear response; instead we have   $r=F^{1/3}$. The key point is that the ``gain'' of the system, $r/F=F^{-2/3}$, diverges at small amplitudes, providing high sensitivity to weak forcings (Fig.~\ref{hopf}). This very high gain does not extend much beyond the resonant frequency $\omega_0$: it drops already by half from its peak when $\vert\omega-\omega_0\vert=3\sqrt{7} F^{2/3}/4$.

\begin{figure}
\begin{center}
\noindent \includegraphics[width=.8\linewidth]{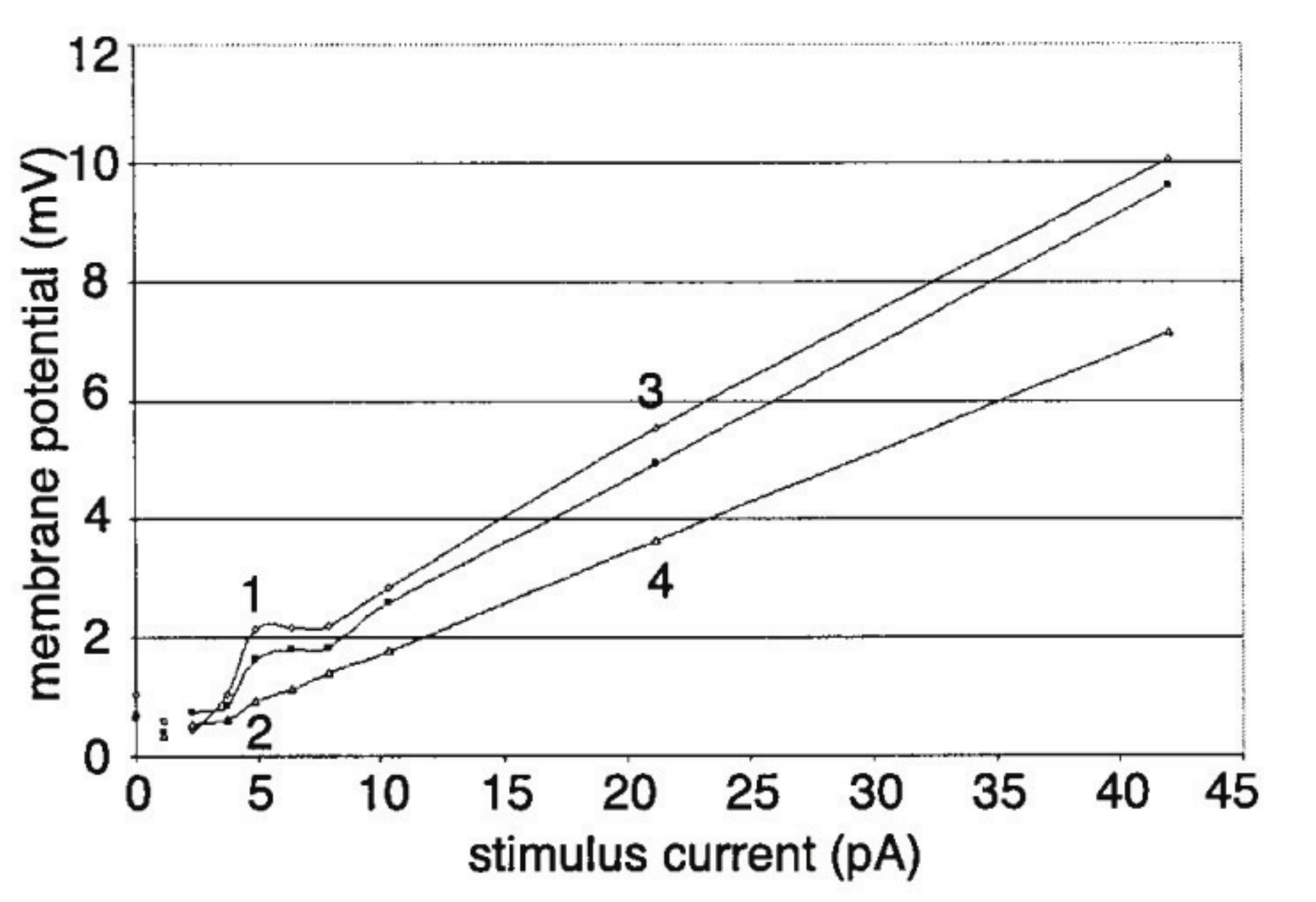}
\end{center}
\caption{Experimental evidence of a Hopf bifurcation in hair cells   \cite{Ospeck:2001p11145}. Shown is the membrane potential as a function of the input current for different input frequencies. Two top curves: an input current oscillating at the resonant frequency (126 Hz) is amplified in a non linear way. Bottom curve: the relation becomes linear when the input frequency is 20 Hz above the resonant frequency.
\label{ampli}
}
\end{figure}

How does this theory fit into what we know about the auditory system? As we have seen, an active process is necessary for amplification. In hair cells, this  active process is provided by hair bundle motility powered by molecular motors, which causes spontaneous oscillations at a characteristic frequency that depends on the geometry of the hair bundle. Hence each cell will be highly selective for one particular frequency. Signal is tranduced by the opening of channels upon deflection of the hair bundle, which has the effect of depolarizing the cell. The interplay of hair bundle motility and external forcing provides the basic ingredients for an excitable Hopf oscillator.  The relevance of Hopf's bifurcation in hair cells was suggested in \cite{Choe:1998p11146}, and its consequences in terms of signal processing was explored in \cite{Eguiluz:2000p11147}. In a parallel effort \cite{Camalet:2000p11141}, an explanation was proposed for how the system tunes itself near the critical point in the oscillating regime ($\mu=0^+$). The idea is that feedback is provided by the activity of the channels themselves, notably through the calcium ion concentration $C$ which controls the activity of the motors responsible for bundle motility. At first approximation one can write $C=C(\mu)$. Channel activity regulates $C$ through 
\beq\label{eq:feedback}
dC/dt=-C/\tau+J(x),
\eeq
where $\tau$ is the relaxation time, $x={\rm Re}(z)$ the hair bundle displacement, and $J(x)$ the displacement-dependent ion flux. For an oscillatory input $x=r\cos(\omega t)$, $\tilde J(r)=\<J(x)\>$ is an increasing function of $r$ (assuming that $J(x)$ is convex). Thus, the non--oscillatory part of $C$ will tune itself at a value such that $C(\mu)=\tau \tilde J(r)=\tau \tilde J(\sqrt{\mu})$.
One can show that, for relevant physical parameters, this drives the system to small values of $\mu$, that is, close to the bifurcation point.

Experiments were able to confirm this picture quantitatively by measuring the voltage response of frog hair cells to an input current. The results showed an enhanced gain for small amplitudes at the resonance frequency (Fig.~\ref{ampli}), as predicted by the theory.  There are also classical experiments in auditory perception that are explained by the Hopf scenario.  In particular,  in the presence of two tones at frequencies $f_1$ and $f_2$, we hear a combination tone at frequency $2f_1 - f_2$, but the apparent intensity of this sound scales linearly with the intensity of the primary tones.  This is completely inconsistent with a system that has a linear response and perturbative nonlinear corrections, but agrees with the $1/3$ power response at the critical point.

Again, we can ask how Hopf bifurcation relates to the equilibrium notion of criticality we have explored before. If we examine the equation governing the evolution of the amplitude $r$ as a function of time, Eq \eqref{eq:hopf}, we can formally rewrite it as the overdamped motion of a  coordinate $r$ in a potential $U$:
\beq\label{eq:landau}
\frac{dr}{dt}=-\frac{\partial U}{\partial r},\quad U(r)=g\frac{r^4}{4}-\mu \frac{r^2}{2}-Fr,
\eeq
with $g=1$. The form of $U$ is familiar: it describes Landau's theory of second order phase transitions. In Landau theory, $\mu$ is a function of the model parameters (notably the temperature), and vanishes at the critical point.    One might object that this dynamical model is not really a many body system, and so can't have a true phase transition.  But in all ears, and especially in the mammalian cochlea, there are many hair cells, tuned to different frequencies, and they are mechanically coupled to one another.    Maximal amplification at each frequency thus requires that the whole system be set such that a macroscopic fraction of the dynamical degrees of freedom are at criticality \cite{Magnasco:2003p11148,Duke:2003p11131}.   Presumably this interacting system should exhibit departures from mean field or Landau behavior, although this has not been explored.

\section{Looking ahead}

We write at a fortunate moment in the development of our subject, when experiments are emerging that hold the promise of connecting decades of theoretical discussion to the real phenomena of life, on many scales.  We hope to have conveyed our reasons for thinking that this is a remarkable development, but also to have conveyed the challenges inherent in this attempt to bring theory and experiment into more meaningful dialogue.  

The first challenge is that we do have somewhat different notions of criticality in different systems, even at the level of theory, and these differences are amplified as we examine the many different approaches to data analysis.  This is a deep problem, not necessarily limited to biological systems.  Except in a few cases, the mathematical language that we use to describe criticality in statistical systems is quite different from the language that we use in dynamical systems.  Efforts to understand, for example, current data on networks of neurons will force us to address the relations between statistical and dynamical criticality more clearly.

The second major challenge is that using the maximum entropy method to analyze real data requires us to solve an inverse statistical mechanics problem.  This problem is tractable far away from critical points, but near criticality it seems very difficult.  If we had more analytic understanding of the problem, it might be possible to identify the signatures of criticality more directly from the measurable correlations, perhaps even allowing us to draw conclusions without explicit construction of the underlying model.  Absent this understanding, there is a serious need for better algorithms.  

A third set of challenges comes from the nature of the data itself.  While we have celebrated the really revolutionary changes in the scale and quality of data now available, there are limitations.  In some cases, such as the flocks of birds, we have relatively few independent samples of the network state; even if we had access to longer time series, the topology of the network is changing as individual birds move through the flock, and we would be forced back to analyzing the system almost snapshot by snapshot.  In other cases, such as protein sequences, we have access to very large data sets but there are unknown biases (the organisms that have been chosen for sequencing).  

A more subtle problem is that, in all cases, the correlations that we observe have multiple origins, some of which are intrinsic to the function of the system and some of which reflect external influences.  For many of the systems we have considered, most of the literature about the analysis of correlations has sought to disentangle these effects, but this work makes clear that it might not be possible to do this without introducing rather detailed model assumptions (e.g., about the mechanisms generating diversity in the antibody repetoire vs. the dynamics of selection in response to antigenic challenge).  In the case of the retina, we know that, quantitatively, roughly half the entropy reduction in the network relative to independent neurons is intrinsic, and half arises in response to the visual stimulus \cite{Schneidman:2006p1273}, but even the ``extrinsic'' correlations are not passively inherited from the outside world, since the strength and form of these correlations depends on the adaptation state of the underlying neural circuitry.    If the networks that we observe, reflecting both intrinsic and extrinsic effects, operate near a critical point, this fact may be more fundamental than the microscopic origins of the correlations.  

Hopefully the discussion thus far has struck the correct balance, exposing the many pieces of evidence pointing toward critical behavior in different systems, but at the same time emphasizing that criticality of biological networks remains a hypothesis whose most compelling tests are yet to come.  To conclude our review, let's take the evidence for criticality at face value, and discuss two questions which are raised by these observations.

The first question is  why biological systems should be nearly critical.  What benefits does operation at this special point in parameters pace provide for these systems?   For birds, we have seen that criticality confers high susceptibility to external perturbations, and this enhanced reactivity endows them with a better defense mechanim against predators. Similarly, in the auditory system, being close to a bifurcation point allows for arbitrarily high gains and accurate frequency selectivity in response weak sounds.   

In neural populations, the naive idea underlying the theory of branching processes makes criticality seem almost inevitable ---\,a middle point between death and epilepsy. However, the function of neural networks is not only to be reactive, but also to carry and process complex information in a collective manner through its patterns of activity.   The observation and analysis of metastable states, both in retinal acitivity analyzed within the maximum entropy framework \cite{Tkacik:2009p7901} and in the activity of cortical slices analyzed with  the theory of branching processes \cite{Haldeman:2005p8046}, suggest that criticality may be coupled to the explosion of these states, allowing for a wider set of coding options.   A more detailed analysis is needed to support this speculation, and to better understand how metastable states can be learned and used in practice for efficient decoding.  More generally, criticality runs counter to simple notions of efficiency in neural coding, suggesting that other principles may be operating, as discussed in Refs  \cite{Tkacik:2006p1289,Tkacik:2009p7901}.   In the case of immune proteins, criticality could be useful for preparedness to attacks, and could result from a tight balance between the expected---prior experience with antigens, as well as hereditary information encoded in the genomic templates---and the unknown.   As in the case of neural coding, the existence of metastable states and their potential for encoding pathogen history may be enhanced by criticality.

The second question is how criticality can be achieved, apparently in so many very different systems.  Critical systems occupy only a thin region (sometimes even a single point) of the parameter space, and it is not clear how biological systems find this region. In some cases, a feedback mechanism can be invoked to explain this adaptation, as in the case of hair cells, where the active process is itself regulated by the amplitude of the oscillations it produces.  In networks of neurons, synaptic plasticity is a good candidate, and there are models that use (more or less) known mechanisms of synaptic dynamics to stabilize a near--critical state  \cite{Magnasco:2009p258102}.  In other cases however, no obvious explanation comes to mind.   

Bird flocks display coherence over very large length scales, which suggests  that the strength of the underlying interactions (that is, the precision with which each bird matches its velocity vector to its neighbors)  is tuned very precisely, but we have no idea how this tuning could be achieved.   In the case of the immune system, feedback does not seem a plausible explanation, because the immune repertoire is constantly renewing itself and out of equilibrium. It is worth noting that a simple mechanism of exponential growth with introduction of random novelty, called the Yule process \cite{Yule:1925p3020}, predicts Zipf's law. However such a model suffers from the same flaws as the branching processes mentioned above: the resulting states are uniformly random, and cannot carry information about their environment, as one would want from an adaptive immune system. Besides, the Yule process does not account for the existence of a constant source of genomic antibody segments. Therefore, the mechanism by which the repertoire maintains criticality remains largely elusive and requires further investigation.

To summarize, we have discussed experimental evidence of criticality in a wide variety of systems, spanning all possible biological scales, from individual proteins to whole populations of animals with high cognitive capacity, in stationary as well as dynamical systems. The wide applicability of the concepts exposed here, fueled by an increasing amount of high quality data, makes for an exciting time.  Ideas which once seemed tremendously speculative are now emerging, independently, from the analysis of real data on many different systems, and the common features seen across so many levels of biological organization encourage us to think that there really are general principles governing the function of these complex systems.

\begin{acknowledgments}
We thank our many collaborators for the pleasure of working together on these ideas:  D Amodei, MJ Berry II, CG Callan, O Marre, M M\'ezard, SE Palmer, R Ranganathan, E Schneidman,  R Segev, GJ Stephens, S Still, G Tka\v{c}ik, and AM Walczak.  In addition, we are grateful to our colleagues who have taken time to explain their own ideas:  A Cavagna, I Giardina, MO Magnasco, and M Weigt.  Speculations, confusions, and errors, of course, remain our fault and not theirs.  This work was supported in part by NSF Grants PHY--0650617 and PHY--0957573, by NIH Grant P50 GM071598, and by the Swartz Foundation; T.~M. was supported in part by the Human Frontiers Science Program.
\end{acknowledgments}

\bibliography{papers_no_url,books}

\end{document}